\documentclass[fleqn,usenatbib,useAMS]{mnras}
\usepackage{natbib}
\usepackage{float} 
\usepackage{graphicx}	
\usepackage{amsmath}	
\usepackage{amssymb}	
\usepackage{multicol}        
\usepackage{bm}		
\usepackage{pdflscape}	
\usepackage{multirow}
\usepackage{comment}
\usepackage[normalem]{ulem} 
\usepackage{xcolor} 

\usepackage[T1]{fontenc}
\usepackage{ae,aecompl}
\usepackage{graphicx}
\usepackage{booktabs}
\usepackage{newtxtext,newtxmath}
\usepackage{natbib}

\title[Guwenbo]{Seeing and Meteorological Analysis at the North-1 and North-2 Points of Muztagh-Ata Site on the Pamir Plateau}

\author[Gu et al.]{
	\large
	\parbox{\textwidth}{
		Wenbo Gu\textsuperscript{1,2}, Ali Esamdin\textsuperscript{1,2}\thanks{E-mail: aliyi@xao.ac.cn}, 
		Cunhai Bai\textsuperscript{1}, Jicheng Zhang\textsuperscript{4,5}, Xuan Zhang\textsuperscript{1}, 
		Guojie Feng\textsuperscript{1}, Letian Wang\textsuperscript{1}, 
		Guangxin Pu\textsuperscript{1}, \\Xinliang Wang\textsuperscript{1},Daiping Zhang\textsuperscript{3}\\[0.5ex]
		\textit{\textsuperscript{1}Xinjiang Astronomical Observatory, Chinese Academy of Sciences, Urumqi 830011, China; guwenbo@xao.ac.cn , aliyi@xao.ac.cn\\
			\textsuperscript{2}School of Astronomy and Space science, University of Chinese Academy of Sciences, Beijing 100049, China \\
			\textsuperscript{3}School of Physics and Technology, Xinjiang University, Urumqi 830046, Xinjiang, China\\
			\textsuperscript{4}School of Physics and Astronomy, Beijing Normal University, Beijing 100875, China\\
			\textsuperscript{5}Institute for Frontiers in Astronomy and Astrophysics, Beijing Normal University, Beijing 100875, China}
	}
}

\date{Accepted 2025 July 21. Received 2025 July 7; in original form 2025 May 31}
\pubyear{2025}

\begin{document}

\label{firstpage}
\pagerange{\pageref{firstpage}--\pageref{lastpage}}
\maketitle

\begin{abstract}
To support the selection of large optical/infrared telescope sites in western China, long-term monitoring of atmospheric conditions and astronomical seeing has been conducted at the Muztagh-Ata site on the Pamir Plateau since 2017. With the monitoring focus gradually shifting northward, three stations were established: the South Point, North-1 point, and North-2 point. The North-1 point, selected as the site for the Muztagh-Ata 1.93 m Synergy Telescope (MOST), has recorded seeing and meteorological parameters since late 2018. In 2023, the North-2 point was established approximately 1.5 km northeast of North-1 point as a candidate location for a future large-aperture telescope. A 10m DIMM tower and a PC-4A environmental monitoring system were installed to evaluate site quality.
This study presents a comparative analysis of data from the North-1 and North-2 points during 2018–2024. The median seeing is $0.89^{\prime\prime}$ at North-1 and $0.78^{\prime\prime}$ at North-2. Both points show clear seasonal and diurnal variations, with winter nights offering optimal observing conditions. 
On average, about 64\% of the nighttime duration per year is suitable for astronomical observations.
Nighttime temperature variation is low: $2.03 ^{\circ}\mathrm{C}$ at North-1 and $2.10 ^{\circ}\mathrm{C}$ at North-2. Median wind speeds are 5–6 m/s, with dominant directions between $210^{\circ}$ and $300^{\circ}$, contributing to stable airflow. Moderate wind suppresses turbulence, while strong shear and rapid fluctuations degrade image quality. These findings confirm that both the North-1 and North-2 points offer high-quality atmospheric conditions and serve as promising sites for future ground-based optical/infrared telescopes in western China.
\end{abstract}




\newpage
\section{Introduction}

\begin{table*}
	\centering
	\caption{Median seeing values measured by French and NIAOT DIMMs at the South site.}
	\label{tab:seeing_south}
	\begin{tabular}{lll}
		\hline
		\textbf{Time Period} & \textbf{Median Seeing (French DIMM)} & \textbf{Median Seeing (NIAOT DIMM)} \\
		\hline
		2017 Jun 23 -- 2017 Nov 15 & 11\,m: 0.79\,\arcsec & Ground: 0.97\,\arcsec \\
		2017 Nov 15 -- 2018 Sep 20 & 11\,m: 0.87\,\arcsec & 6\,m: 0.87\,\arcsec \\
		2018 Sep 21 -- 2018 Nov 20 & 11\,m: 0.71\,\arcsec & 11\,m: 0.72\,\arcsec \\
		\hline
	\end{tabular}
\end{table*}

The Muztagh-Ata site is located on the Pamir Plateau in southwestern China, at approximately \(38^{\circ}21^{\prime}\mathrm{N},\,74^{\circ}54^{\prime}\mathrm{E}\), with an elevation of around 4500\,m. This high-altitude region features extremely low annual precipitation and minimal human activity, offering a naturally pristine environment for astronomical observations. Since January 2017, the site has undergone continuous and comprehensive monitoring for eight consecutive years. Substantial progress has been made in evaluating atmospheric and meteorological conditions~\citep{2020RAA2086X,2024MNRAS.535.3543Z}, observational continuity~\citep{2023RAA....23d5015X,2024RAA....24c5003G}, night-sky brightness~\citep{2020RAA2086X}, seeing~\citep{2020RAA....20...87X,2024RAA....24a5006W,2025MNRAS.539.2077Z}, atmospheric water vapor content~\citep{Xu2022}, and near-surface turbulence characteristics~\citep{2024MNRAS.535.1193G}. These results collectively demonstrate that the site offers excellent conditions for optical astronomical observations.

\begin{figure}
	\includegraphics[width=\columnwidth]{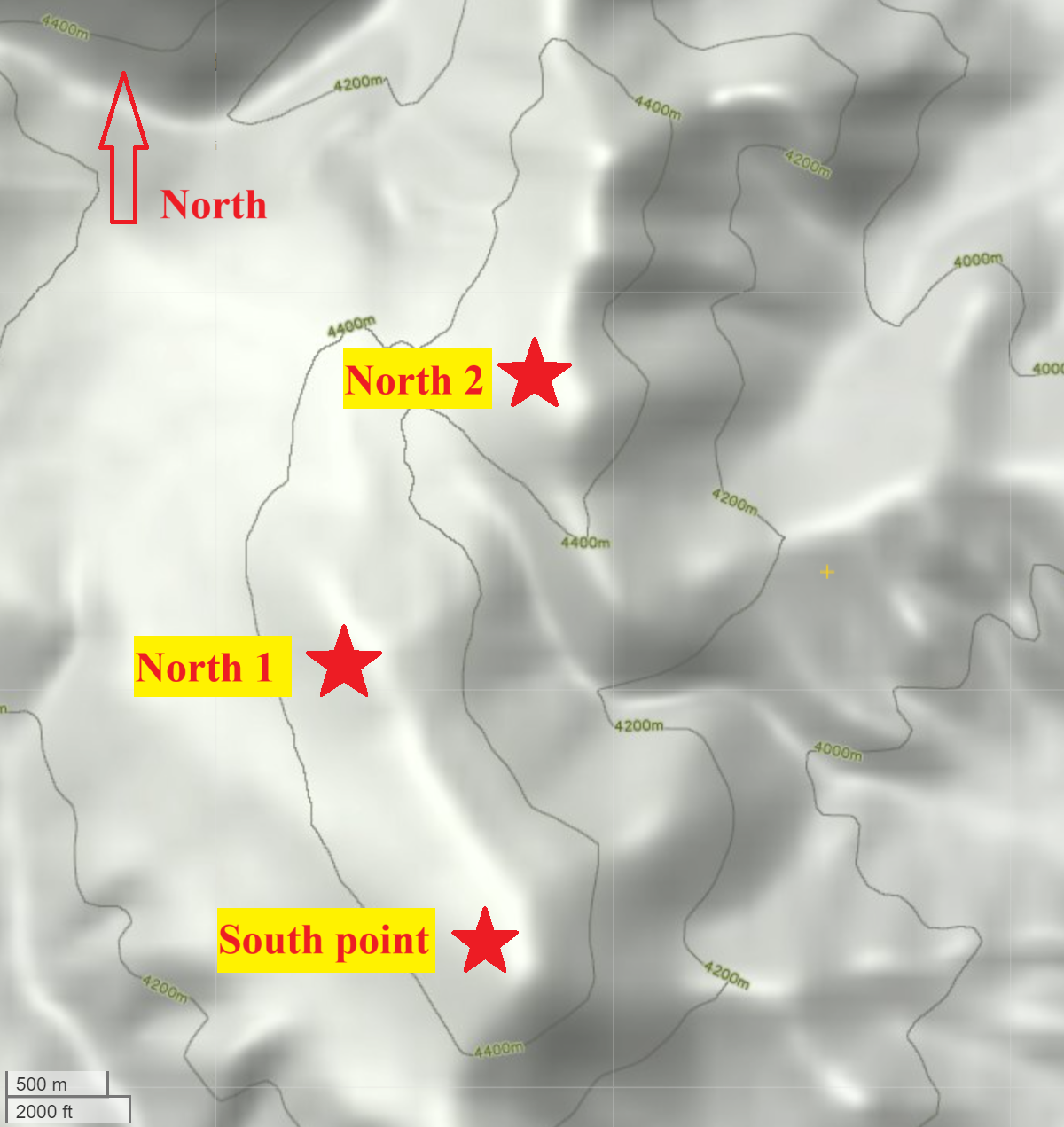}
	\caption{Topographic contour map of the Muztagh-Ata site on the Pamir Plateau, showing the locations of the three monitoring points: the South Point, North-1 Point, and North-2 Point.}
	\label{fig:muztaghnorth2}
\end{figure}

Since the launch of the monitoring program in 2017, three independent observation points have been established at the Muztagh-Ata site: the South Point, North-1 Point, and North-2 Point. The South Point was the first to become operational in 2017 and remained active until March 2020. During this period, the primary observational focus gradually shifted northward. North-1 Point has been in operation since August 2017 and currently serves as the construction site of the Muztagh-Ata 1.93 m Synergy Telescope (MOST).
Building upon this foundation, and in preparation for the site selection of China’s future large-aperture optical/infrared telescope, North-2 Point was established in November 2023 to evaluate its potential to meet the stringent optical and environmental criteria required for such a facility. Monitoring at North-2 Point concluded in September 2024. The site is located approximately 1.5\,km northeast of North-1 Point. The relative spatial layout of the three observation points is shown in Figure~\ref{fig:muztaghnorth2}.

Atmospheric optical turbulence along the line of sight degrades the quality of astronomical observations, with seeing being a key parameter that quantifies the degree of stellar image blurring induced by such turbulence. Seeing is also one of the most critical indicators for assessing the observational quality of the Muztagh-Ata site. The Differential Image Motion Monitor (DIMM) is the standard instrument used to measure this parameter~\citep{sarazin1990eso}. The seeing obtained from DIMM reflects the integrated turbulence effects along the entire optical path. Its high-precision measurements not only provide essential references for the design and construction of large-aperture telescopes, but also offer critical support for adaptive optics (AO) systems~\citep{vernin1995measuring,2002PASP..114.1156T} and other advanced observing techniques.
In earlier studies, seeing measurements at the South Point of the Muztagh-Ata site were conducted using two instruments: a French DIMM and a NIAOT DIMM, installed at different heights. Both DIMM systems applied zero-exposure time correction. A comparison of the results obtained from these two instruments is shown in Table~\ref{tab:seeing_south}~\citep{2020RAA....20...87X}.

At the North-1 Point, seeing monitoring commenced in 2018 with the deployment of a Differential Image Motion Monitor (DIMM) system developed by the Nanjing Institute of Astronomical Optics and Technology (NIAOT), Chinese Academy of Sciences, mounted on a 6m tower. In 2021, a 30m meteorological tower was constructed, equipped with five ultrasonic anemometers, five temperature sensors, and dedicated sensors for relative humidity and atmospheric pressure, enabling continuous meteorological observations. Additionally, a Sky Quality Meter (SQM) photometer and a visible-light all-sky camera were installed to monitor night-sky brightness and cloud coverage.
Using DIMM data from the 6\,m tower at North-1 Point between 2018 and 2022, \citet{2025MNRAS.539.2077Z} conducted a systematic analysis of seeing conditions and their correlation with the 200\,hPa wind speed (V200), reporting a median seeing of 0.89\arcsec. Building on this foundation, DIMM observations have been continuously carried out at the site.

In December 2023, a 10\,m-high seeing tower was constructed at North-2 Point, where a French DIMM system and a PC-4A meteorological monitoring system were installed to monitor seeing and five key meteorological parameters. DIMM observations at North-2 concluded on 10 June 2024. Based on data collected from December 2018 to the present at both North-1 and North-2—including temperature, wind speed, atmospheric pressure, wind direction, and humidity, along with seeing measurements from the 6\,m tower at North-1 and the 10\,m tower at North-2—we performed a comprehensive analysis of the seeing and meteorological characteristics at both points.

This paper focuses on the on-site monitoring campaigns conducted at the North-1 and North-2 Points of the Muztagh-Ata site since December 2018, aiming to provide essential references for the future site selection of large optical telescopes. Section~2 outlines the working principles of the DIMM system, the instruments used for seeing and meteorological measurements, and the procedures for data acquisition and quality control. 
Section~3 summarizes the monthly and hourly seeing statistics, key meteorological parameters, and their relation to seeing—especially wind shear. It also includes an 8-year analysis of observable time and its monthly distribution.
Section~4 summarizes the main findings of this study and outlines directions for future research.

\section{Data and Methods}
\subsection{Measurements}
Seeing (\(\varepsilon\)) is typically defined by measuring the full width at half maximum (FWHM) of a long-exposure stellar image obtained under zenith observation and a wavelength of 500\,nm \citep{pardo2002modeling}. Atmospheric turbulence studies generally rely on seeing (\(\varepsilon\)) measurements.

\begin{figure}
	\includegraphics[width=\columnwidth]{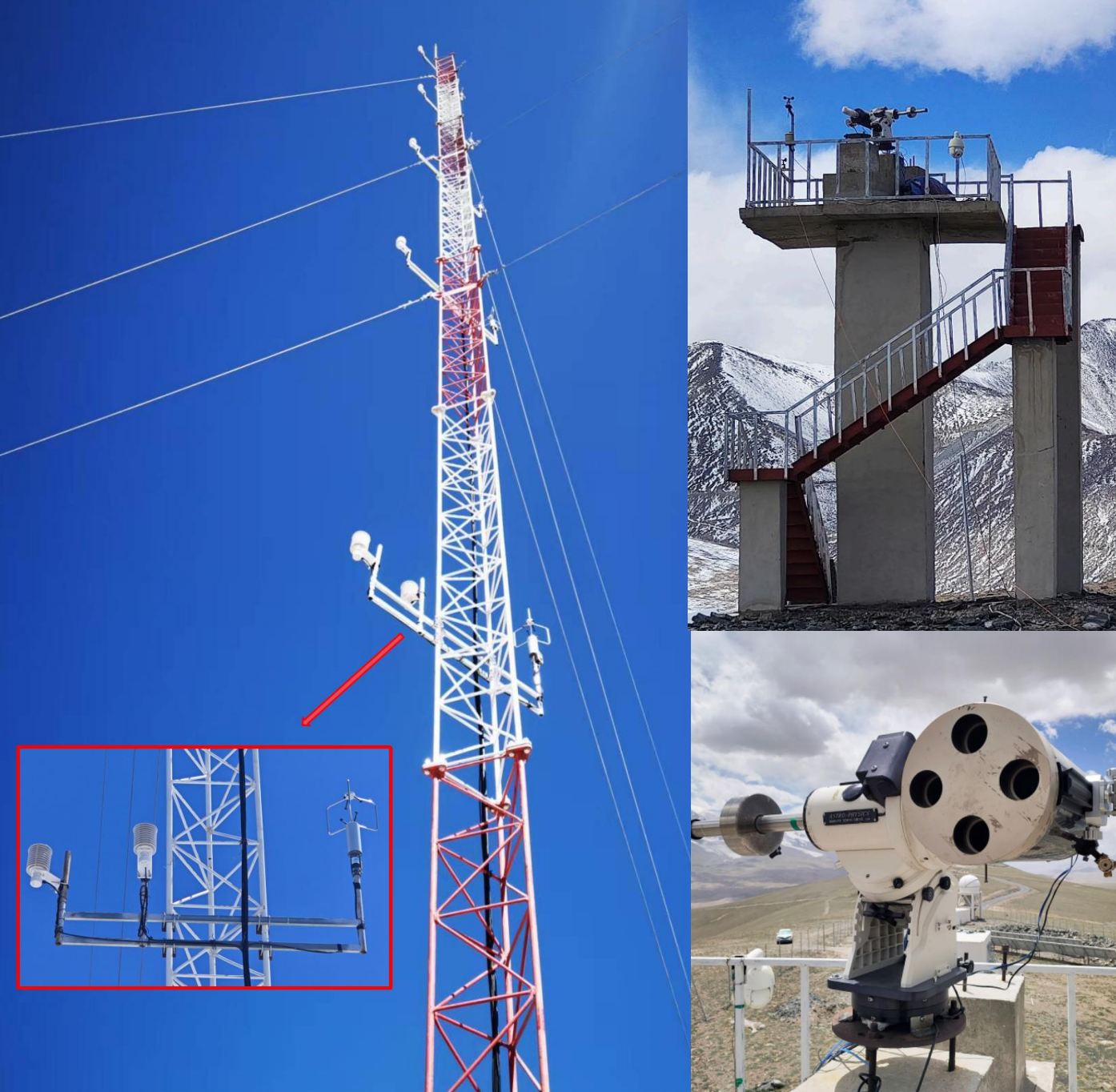}
	\caption{Left: the 30-meter meteorological tower at  the North-1 point. Right: the NIAOT DIMM and 6m tower.}
	\label{fig:muztaghnorthfig2_1}
\end{figure}

The relationship between seeing and optical turbulence intensity can be written as:
\begin{equation}
\epsilon = 0.98 \frac{\lambda}{r_{0}} = 5.25 \lambda^{-1/5} \left[ \int_{0}^{\infty} C_{n}^{2}(h) dh \right]^{3/5}
\end{equation}
where \(\lambda\) is the observing wavelength, \(\int_0^{\infty} C_{n}^{2}(h)\,dh\) represents the optical turbulence energy distribution, $r_{0}$ is the Fried parameter, \(C_{n}^{2}\) is the refractive index structure constant, and \(h\) denotes altitude. For DIMM measurements, the central wavelength is set to \(\lambda = 0.5\,\mu\mathrm{m}\), and the final results are converted to the zenith angle. Here, \(\varepsilon\) is the seeing measured by DIMM. The DIMM system determines the stellar centroid positions by capturing short-exposure images, with the exposure time automatically adjusted based on the star’s magnitude.
At the North-1 Point of Muztagh-Ata site, the DIMM tower stands 6 meters tall and adopts an open-frame design without a protective dome. It is equipped with a NIAOT DIMM system dedicated to seeing measurements. Adjacent to it stands a 30-meter meteorological tower equipped with five ultrasonic anemometers (Young 81000), five temperature sensors (Young 41342), a barometric pressure sensor (Young 61302V), and a relative humidity sensor (Young 41382VC). Figure~\ref{fig:muztaghnorthfig2_1} shows the on-site images of these devices. The 30m tower is on the left, and the dimm device is on the right.

\begin{figure}
	\includegraphics[width=\columnwidth]{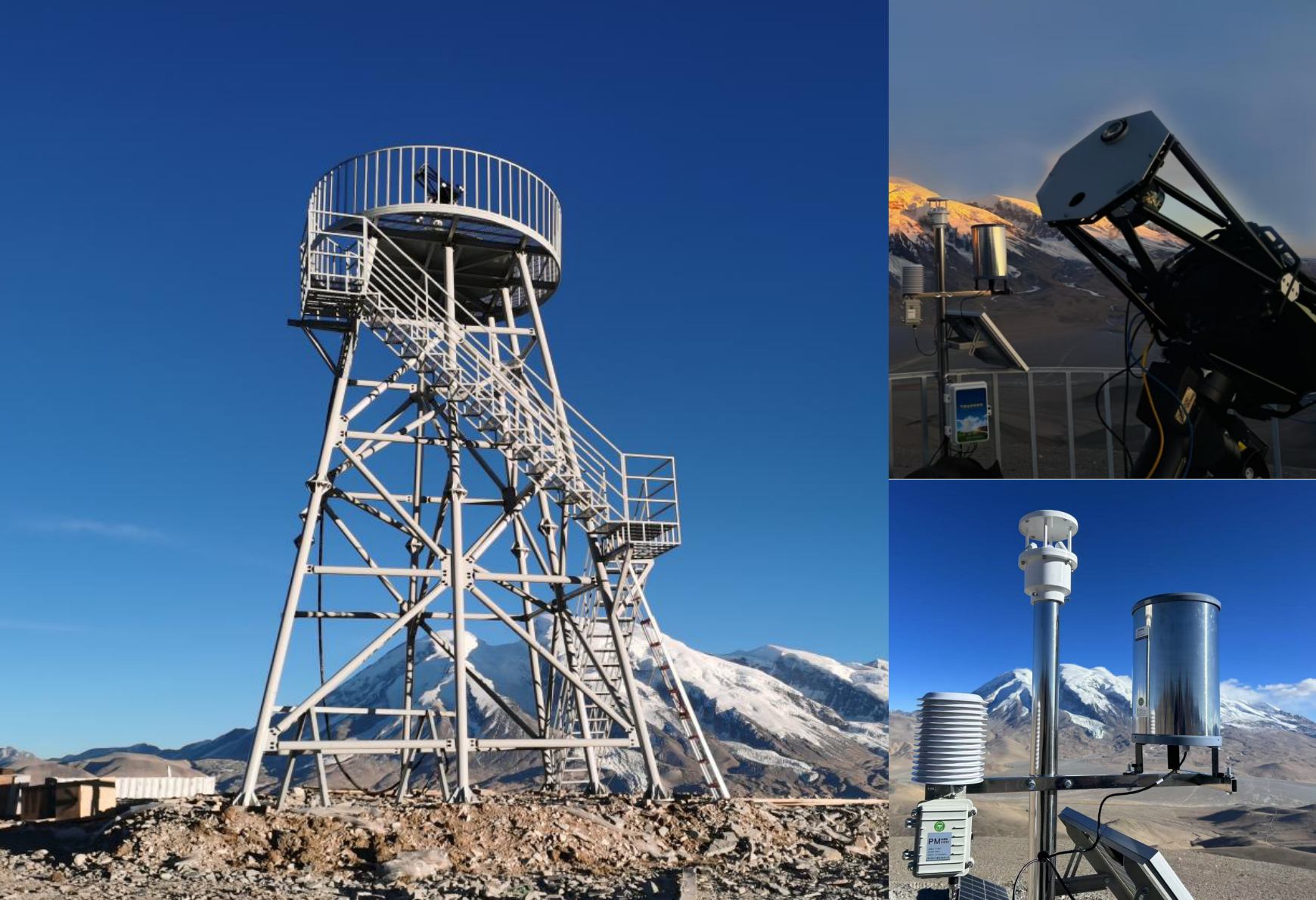}
	\caption{Left: the 10m tower at North-2 site. Right: the French DIMM and PC-4A environmental monitoring instrument.}
	\label{fig:muztaghnorthfig2}
\end{figure}

At the North-2 Point of  Muztagh-Ata site, the DIMM tower is 10\,m tall and features an open-frame structure without a dome. The platform is equipped with a French DIMM system for seeing measurements and a PC-4A environmental monitoring unit for real-time meteorological data acquisition. As shown in Figure~\ref{fig:muztaghnorthfig2}, the 10\,m DIMM tower appears on the left, while the French DIMM and the PC-4A instruments are shown on the right.

As shown in Table~\ref{tab:dimm_specs}, the French DIMM system, provided by Alcor System, uses a 300\,mm f/8 Ritchey–Chrétien telescope with a focal length of 2400\,mm, two 51\,mm sub-apertures spaced 240\,mm apart, and a DMK 33GX174 CCD camera with automatic exposure from 0.5 to 1000\,ms. Observations are conducted at 550\,nm, and one seeing value is produced every 1000 frames (typically every 5--8 seconds). It was installed on a 10\,m tower at the North-2 Point in December 2023 and operated only on cloud-free nights.
The NIAOT DIMM, developed by the Nanjing Institute of Astronomical Optics and Technology, employs a 200\,mm f/8 telescope with a focal length of 1600\,mm, four 50\,mm sub-apertures spaced 149\,mm apart, and a Basler aca2040 CMOS camera (2048\,$\times$\,2048 pixels, 5.5\,$\mu$m). It uses fixed alternating exposures of 5\,ms and 10\,ms at 500\,nm and produces one seeing value per minute. Observations were conducted on a 6\,m tower at the North-1 Point from 2018 to 2023. 
The measurements from both DIMM systems were corrected for zero exposure time and converted to zenith direction.

Meteorological parameters at the North-1 Point of the Muztagh-Ata site are measured by instruments mounted on a 30-meter meteorological tower, which has been operational since 30 September 2021. The Young 81000 ultrasonic anemometers measure wind speed and direction at a sampling frequency of 20\,Hz (i.e., 20 measurements per second), while the Young 41342 temperature sensor, Young 61302V barometric pressure sensor, and Young 41382VC relative humidity sensor each record one data point every 30 seconds.
To standardize the temporal resolution and facilitate subsequent statistical analysis, all sensor data from the 30-meter tower were averaged over 1-minute intervals to produce the final dataset. As of 17 June 2024, a total of 965 valid observing days had been recorded at North-1, resulting in approximately 1,354,086 meteorological data entries.

At the North-2 Point, the PC-4A meteorological instrument has been operational since 22 January 2024. It continuously samples five key meteorological parameters at a fixed frequency of 1\,Hz, generating one data record per second. By 17 September 2024, 180 valid observing days had been recorded, yielding approximately 247,224 meteorological data entries. The measurement range, resolution, and accuracy of each sensor are summarized in Table~\ref{tab:PC4A_specs}.

All time references in this paper are expressed in Coordinated Universal Time (UTC). Daytime and nighttime are defined based on astronomical twilight.

\begin{table*}
	\centering
	\caption{Instrument specifications of the French DIMM and NIAOT DIMM systems.}
	\label{tab:dimm_specs}
	\begin{tabular}{lll}
		\hline
		\textbf{Parameter} & \textbf{French DIMM} & \textbf{NIAOT DIMM} \\
		\hline
		Telescope Aperture (mm)         & 300                             & 200 \\
		Focal Ratio                     & f/8                             & f/8 \\
		Focal Length (mm)              & 2400                            & 1600 \\
		Number of Sub-apertures        & 2                               & 4 \\
		Sub-aperture Diameter (mm)     & 51                              & 50 \\
		Sub-aperture Separation (mm)   & 240                             & 149 \\
		Detector / CCD Camera          & DMK 33GX 174                    & Basler aca2040, 2k × 2k (5.5\,$\mu$m) \\
		Exposure Mode                  & Auto-adjusted (interlaced)      & Fixed alternating (interlaced) \\
		Exposure Time (ms)             & 0.5–1000 (auto-adjusted)        & Alternating between 5 and 10 ms \\
		Wavelength (nm)                & 550                             & 500 \\
		Output Frequency               & 1 seeing value per 1000 images  & 1 seeing value per minute \\
		Zero-Exposure Time Correction  & Corrected to zero exposure time & Corrected to zero exposure time \\
		Zenith Correction & Applied & Applied \\

		\hline
	\end{tabular}
\end{table*}

\begin{table*}
	\centering
	\caption{Specifications of meteorological and environmental sensors used at the Muztagh-Ata site.}
	\label{tab:PC4A_specs}
	\begin{tabular}{lllll}
		\hline
		\textbf{Sensor} & \textbf{Parameter} & \textbf{Measurement Range} & \textbf{Resolution} & \textbf{Accuracy} \\
		\hline
		\multirow{3}{*}{Young 81000} 
		& Wind speed & 0--40\,m\,s$^{-1}$ & 0.01\,m\,s$^{-1}$ & $\pm$1\% (0--30\,m\,s$^{-1}$), $\pm$3\% (30--40\,m\,s$^{-1}$) \\
		& Wind direction & 0--359.9$^\circ$ & 0.1$^\circ$ & $\pm$2$^\circ$ (1--30\,m\,s$^{-1}$), $\pm$5$^\circ$ (30--40\,m\,s$^{-1}$) \\
		& Sonic temperature & --50 to +50$^\circ$C & 0.01$^\circ$C & $\pm$2$^\circ$C \\
		\hline
		Young 41342 
		& Temperature & --50 to +50$^\circ$C & 0.1$^\circ$C & $\pm$0.3$^\circ$C \\
		\hline
		\multirow{1}{*}{Young 61302V}  & pressure & 500--1100\,hPa & 0.2\,hPa & $\pm$0.2\,hPa\\
		\hline
		\multirow{2}{*}{Young 41382VC} 
		& Relative humidity & 0--100\% RH & 0.1\% RH & $\pm$1\% RH \\
		& Temperature & --50 to +50$^\circ$C & 0.1$^\circ$C & $\pm$0.3$^\circ$C \\
		\hline
		\multirow{6}{*}{PC-4A} 
		& Temperature & --50 to +80$^\circ$C & 0.1$^\circ$C & $\pm$0.1$^\circ$C \\
		& Relative humidity & 0--100\% RH & 0.1\% RH & $\pm$2\% ($\leq$80\%), $\pm$5\% ($>$80\%) \\
		& Dew point temperature & --40 to +50$^\circ$C & 0.1$^\circ$C & $\pm$0.2$^\circ$C \\
		& Wind direction & 0--360$^\circ$ & 3$^\circ$ & $\pm$3$^\circ$ \\
		& wind speed & 0--70\,m\,s$^{-1}$ & 0.1\,m\,s$^{-1}$ & $\pm$(0.3 + 0.03V)\,m\,s$^{-1}$ \\
		& Pressure & 550--1060\,hPa & 0.1\,hPa & $\pm$0.3\,hPa \\
		\hline
	\end{tabular}
\end{table*}

\subsection{Seeing Data Preprocessing}

During continuous DIMM monitoring at the Muztagh-Ata site, it was observed that the passage of clouds across the line of sight caused stellar images to dim or flicker abruptly, resulting in distortions in the measured differential image motion. Consequently, the derived seeing values were often contaminated by noise, leading to spurious and non-physical fluctuations. If such cloud-contaminated data—representing so-called “false turbulence”—are included in statistical analyses, they can introduce systematic biases and obscure the true distribution of atmospheric seeing. Therefore, identifying and removing cloud-affected periods during the data processing stage is essential to ensure the reliability of seeing statistics.

\begin{figure}
	\includegraphics[width=\columnwidth]{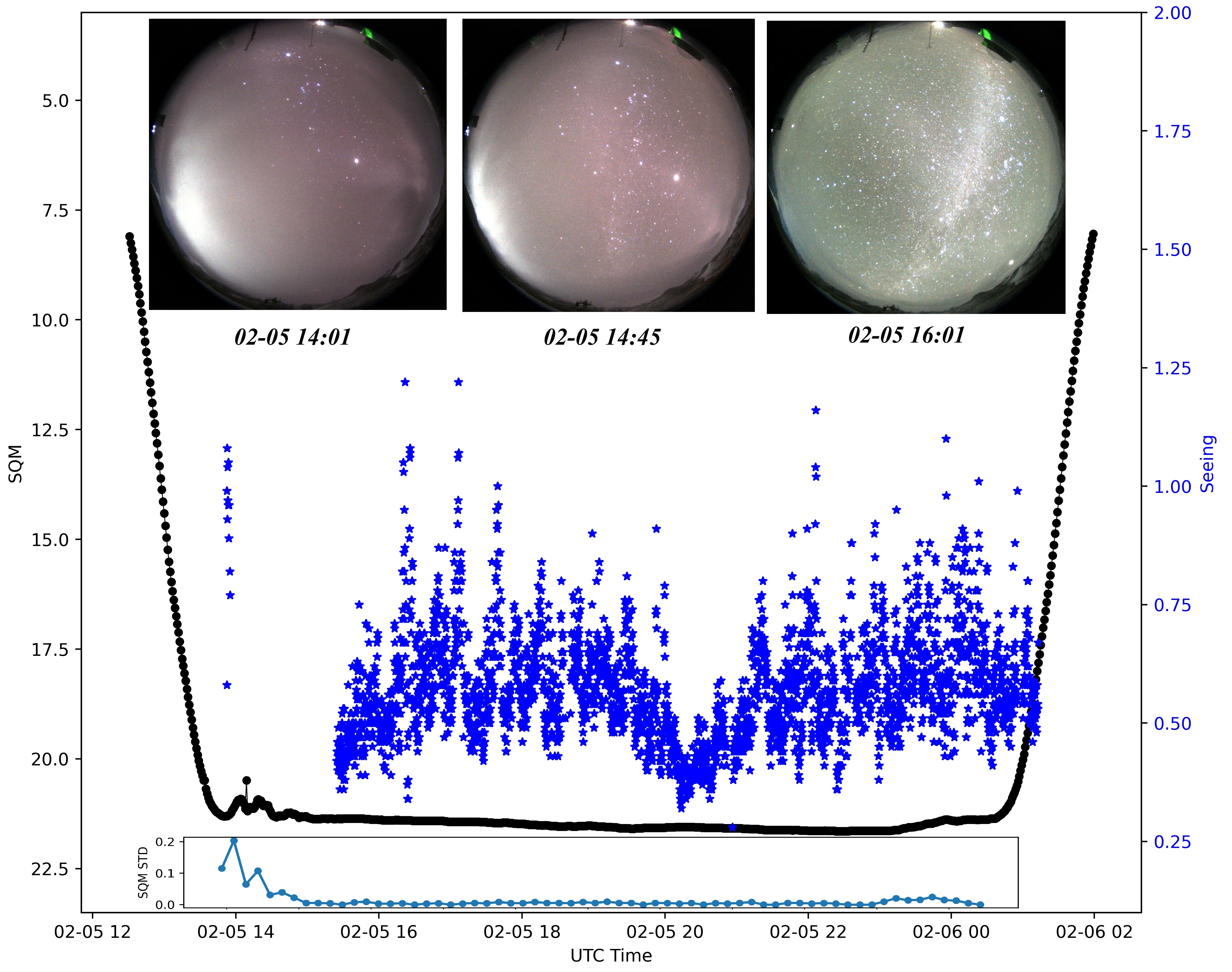}
	\caption{SQM readings and DIMM seeing values at the Muztagh-Ata site on 5 February 2024. The inset shows the temporal variation of the SQM standard deviation, along with selected all-sky camera images corresponding to key time intervals.}
	\label{fig:sqmcloud}
\end{figure}

To address this issue, we adopted a method proposed by \citet{2020MNRAS.493.2463C}, which identifies cloud passages based on the standard deviation of Sky Quality Meter (SQM) readings. This approach has also been referenced and applied at the Lenghu site~\citep{deng2021lenghu}. As shown in Figure~\ref{fig:sqmcloud}, we present a representative case from the Muztagh-Ata site, including the full-night SQM time series, corresponding DIMM seeing data, and selected all-sky images.
A significant cloud event occurred between 14:01 and 14:45 UTC on 5 February, during which the standard deviation of SQM readings exhibited notable fluctuations. After 15:00, the sky became clear again, and SQM values remained low and stable.

To assess the impact of cloud contamination during seeing monitoring, we applied a statistical filtering method based on the standard deviation of SQM readings, with verification provided by all-sky camera images. The filtering procedure comprises the following steps:

\begin{enumerate}
	\item \textbf{Baseline Determination under Clear and Moonless Conditions}
	
	Time windows were selected in which all-sky images clearly showed cloud-free skies, and the Moon was either in the new moon phase or positioned below --6° altitude. For these intervals, the 10-minute standard deviation of the SQM time series was calculated. It was found that 95\% of these baseline data points exhibited standard deviations below 0.030\,mag\,arcsec$^{-2}$. Therefore, we adopted $\sigma_{\mathrm{baseline}} = 0.030$\,mag\,arcsec$^{-2}$ as the threshold for cloud-free, moonless conditions.
	
	\item \textbf{Lunar Phase Correction}
	
	To account for sky brightness variability due to lunar phase, we examined SQM standard deviations under both new moon and full moon conditions. Results showed that the threshold $\sigma$ increases under bright moonlight. Based on iterative testing, we introduced a sinusoidal correction to dynamically modulate the threshold as a function of lunar age:
	\begin{equation}
		\sigma(d) = \sigma_{\mathrm{baseline}} + \Delta \left| \sin\left( \frac{\pi d}{T} \right) \right|, \quad \Delta = 0.023,\quad T = 29.53\,\mathrm{d},
	\end{equation}
	where $d$ is the number of days since the last new moon, $\Delta$ is the empirical modulation amplitude, and $T$ is the lunar cycle period.
	
	\item \textbf{Cloud Detection and All-Sky Camera Validation}
	
	If the SQM standard deviation in a 10-minute window exceeded the threshold $\sigma(d)$, the interval was flagged as a “cloud candidate.” These candidates were manually validated using all-sky images taken at corresponding times. If persistent cloud patterns were visible, the interval was confirmed as “cloud-affected”; otherwise, it was treated as SQM noise and retained.
	
	\item \textbf{DIMM Data Filtering}
	
	All time intervals confirmed to be affected by clouds were excluded from the DIMM seeing time series to ensure the accuracy and statistical reliability of the results.
\end{enumerate}

Following the procedure described above, we applied the cloud-filtering algorithm to the full DIMM seeing dataset, excluding all time intervals suspected to be affected by cloud passages. This ensures that the retained DIMM data reflect true atmospheric seeing conditions, free from cloud contamination.

\section{Result}

\subsection{Seeing Statistics and Analysis}
At the North-1 Point of Muztagh-Ata site , seeing monitoring began in 2018 using a NIAOT DIMM system and continued until 2023. Throughout the entire observation period, the instrument was mounted on a 6m tower. A total of 297 valid nights were recorded, yielding 73,499 effective seeing measurements.
At the North-2 Point, seeing monitoring started in late 2023 using a French DIMM system and continued through 2024. During the observation period, the instrument was installed on a 10m tower. A total of 122 valid nights were obtained, resulting in 217,527 seeing measurements that effectively reflect the atmospheric quality of the site.

\begin{figure}
	\includegraphics[width=\columnwidth]{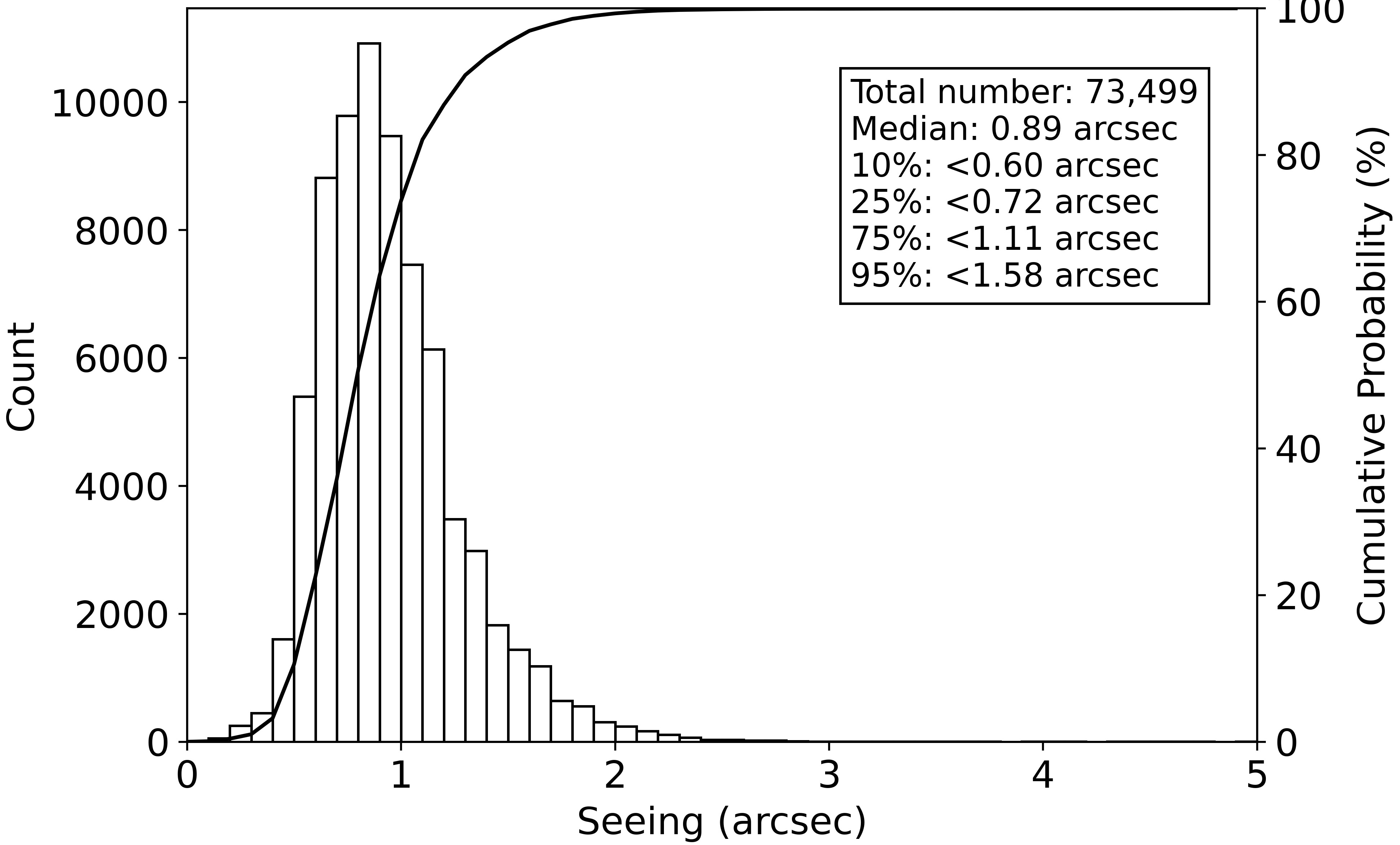}
	\caption{Histogram and cumulative distribution function of seeing measurements at the North-1 point of the Muztagh-Ata site, based on observations obtained from 2018 to 2023 using the NIAOT DIMM mounted on a 6\,m tower.}
	\label{fig:nanjing_hist_cdf}
\end{figure}

\begin{figure}
	\includegraphics[width=\columnwidth]{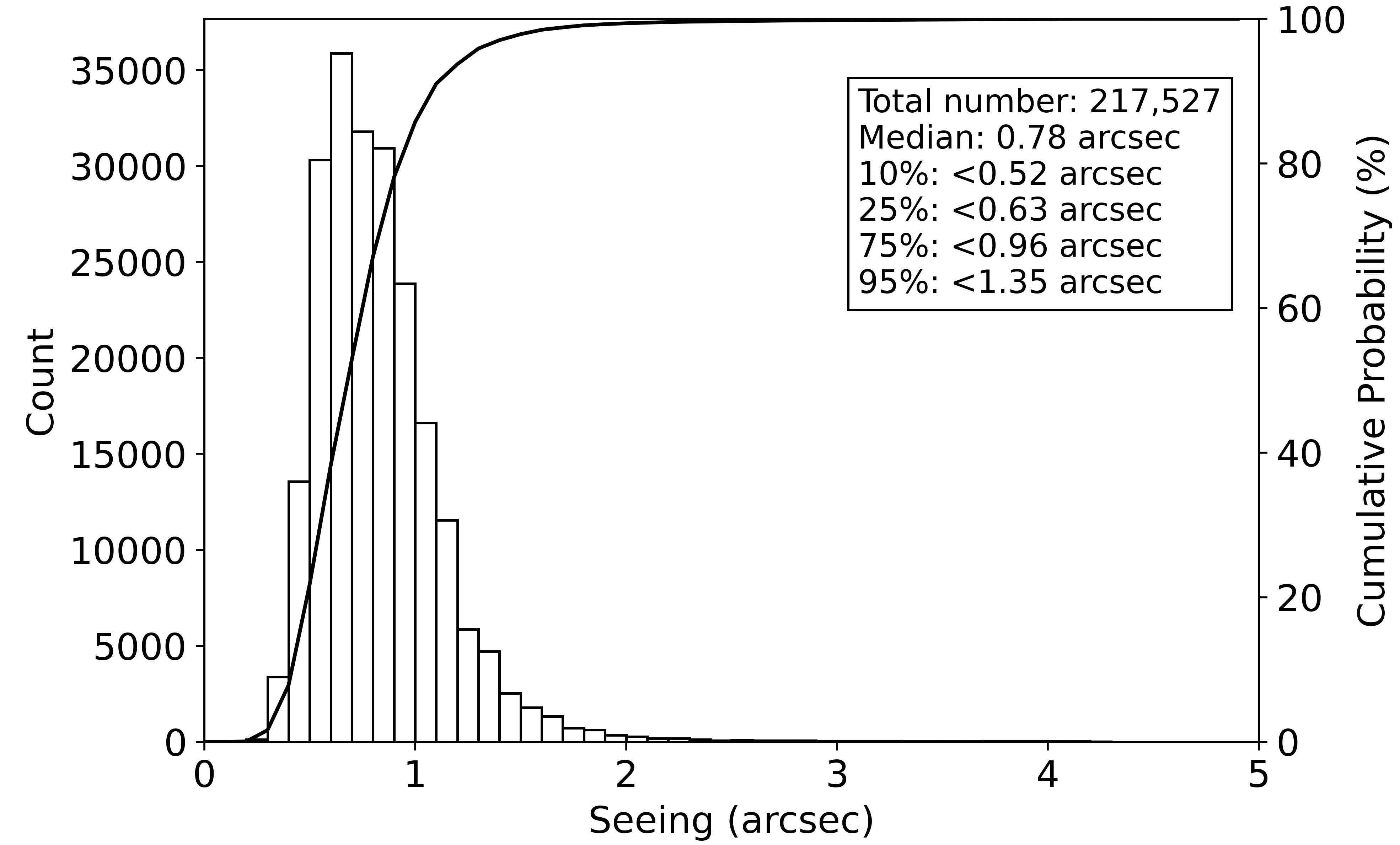}
	\caption{Histogram and cumulative distribution function of seeing measurements at the North-2 point of the Muztagh-Ata site, obtained from 14 December 2023 to 15 June 2024 using the French DIMM system mounted on a 10\,m tower.}
	\label{fig:french_hist_cdf}
\end{figure}

Figure~\ref{fig:nanjing_hist_cdf} and Figure~\ref{fig:french_hist_cdf} show the statistical and cumulative probability distributions of the seeing values measured by the NIAOT and French DIMM systems, respectively. At North-1, the median seeing was 0.89\,arcseconds, with 75\% of values below 1.11\,arcseconds. At North-2, the median seeing was 0.78\,arcseconds, with 75\% of values below 0.96\,arcseconds.

Table~\ref{tab:seeing_comparison} provides a summary of the seeing conditions at the Muztagh-Ata North-1 and North-2 Points in comparison with several internationally recognized observatory sites. These include Mauna Kea in Hawaii, USA~\citep{schock2009thirty}; La Palma in the Canary Islands, Spain~\citep{ma2020night}; Cerro Paranal and Cerro Chico in the Atacama Desert, Chile~\citep{giovanelli2001optical}; and Dome C in Antarctica~\citep{aristidi2009dome}. Both North-1 and North-2 demonstrate competitive median seeing values. In particular, North-2 shows stable and high-quality seeing, with a median of 0.78\,arcseconds and an interquartile range well below 1\,arcsecond. North-1, despite a slightly higher median, maintains consistently good conditions as well.
These results indicate that both points at the Muztagh-Ata site offer favorable optical observing conditions. The combination of low median seeing and atmospheric stability highlights their strong potential as candidate locations for future large-aperture ground-based telescopes.

\begin{table}
	\centering
	\caption{Comparison of seeing conditions at the Muztagh-Ata North-1 and North-2 points with other internationally recognized observatory sites. The table includes DIMM tower heights and the 25th percentile, median, and 75th percentile seeing values.}
	\label{tab:seeing_comparison}
	\renewcommand{\arraystretch}{1.0} 
	\setlength{\tabcolsep}{3pt} 
	\begin{tabular}{l c c c c}
		\hline
		\textbf{Site} & \textbf{DIMM Tower Height} & \textbf{25\%} & \textbf{Median} & \textbf{75\%} \\
		\hline
		Muztagh-Ata North-1 & 6m & 0.72 & 0.89 & 1.11 \\
		Muztagh-Ata North-2 & 10m & 0.63 & 0.78 & 0.97 \\
		Mauna Kea & 7m & 0.57 & 0.75 & 1.03 \\
		La Palma & 5m & 0.62 & 0.80 & 1.06 \\
		Cerro Paranal & 6m & 0.71 & 0.80 & 0.87 \\
		Cerro Chico & 2.5m & 0.66 & 0.71 & 0.87 \\
		Dome C & 20m & 0.43 & 0.84 & 1.55 \\
		Dome C (FA) & 20m & - & 0.30 (15.7\%) & - \\
		\hline
	\end{tabular}
\end{table}

\begin{table*}
	\centering
	\caption{Monthly median and quartile seeing values measured by the NIAOT DIMM at the North-1 Point (6m tower) from 2017 to 2023. A dash (–) denotes months with no available data.}
	\label{tab:monthly_stats}
	\begin{tabular}{l *{4}{c} *{4}{c} *{4}{c}}
		\toprule
		Month 
		& \multicolumn{4}{c}{2018}
		& \multicolumn{4}{c}{2019}
		& \multicolumn{4}{c}{2020} \\
		\cmidrule(lr){2-5} \cmidrule(lr){6-9} \cmidrule(lr){10-13}
		& Count & Median & 25\% & 75\%
		& Count & Median & 25\% & 75\%
		& Count & Median & 25\% & 75\% \\
		\midrule
		Jan   & -- & --   & --   & --   & -- & -- & -- & -- & 4407 & 0.98   & 0.79  & 1.19\\
		Feb   & -- & --   & --   & --   & -- & -- & -- & -- & -- & --   & --   & --   \\
		Mar   & -- & --   & --   & --   & -- & -- & -- & -- & -- & --   & --   & --   \\
		Apr   & -- & --   & --   & --   & 673 & 0.83 & 0.70 & 0.95 & -- & -- & -- & -- \\
		May   & -- & --   & --   & --   & 2326 & 0.81 & 0.68 & 0.96 & --& -- & -- & -- \\
		Jun   & -- & --   & --   & --   & 545 & 0.77   & 0.66  & 1.13 & -- & -- & -- & -- \\
		Jul   & -- & --   & --   & --   & 3807 & 0.8  & 0.68 & 0.94 & 1553 & 0.82 & 0.68 & 0.99 \\
		Aug   & -- & --   & --   & --   & 2468 & 0.88 & 0.76 & 1.08 & -- & -- & -- & -- \\
		Sep   & -- & --   & --   & --   & 9876 & 0.86 & 0.73 & 1.05 & 5645 & 0.99 & 0.86 & 1.18 \\
		Oct   & -- & --   & --   & --   & 5523 & 0.71 & 0.57 & 0.86 & 3168 & 0.84 & 0.72 & 1.05 \\
		Nov   & -- & --   & --   & --   & 5182 & 0.73 & 0.62 & 0.90 & -- & -- & -- & -- \\
		Dec   & 2940 & 0.68 & 0.52 & 1.06 & 365 & 0.57 & 0.49 & 0.69 & --& -- & -- & -- \\
		\midrule
		Month 
		& \multicolumn{4}{c}{2021}
		& \multicolumn{4}{c}{2022}
		& \multicolumn{4}{c}{2023} \\
		\cmidrule(lr){2-5} \cmidrule(lr){6-9} \cmidrule(lr){10-13}
		& Count & Median & 25\% & 75\%
		& Count & Median & 25\% & 75\%
		& Count & Median & 25\% & 75\% \\
		\midrule
		Jan   & 2485 & 1.07 & 0.93 & 1.28  & 1054 & 0.95 & 0.80 & 1.33  & -- & -- & -- & -- \\
		Feb   & -- & --   & --   & --      & 1153 & 0.83 & 0.73 & 0.98  & 24 & 0.89 & 0.78 & 1.02 \\
		Mar   & -- & --   & --   & --      & 878  & 1.12 & 0.96 & 1.31  & -- & -- & -- & -- \\
		Apr   & 1754 & 1.06 & 0.89 & 1.22  & 2356 & 0.92 & 0.77 & 1.10  & -- & -- & -- & -- \\
		May   & -- & -- & -- & --          & 12  & 1.13 & 0.96 & 1.27   & -- & -- & -- & -- \\
		Jun   & -- & --   & --   & --      & 93  & 0.89 & 0.82 & 0.97   & -- & -- & -- & -- \\
		Jul   & 973 & 1.00 & 0.88 & 1.10   & 504 & 1.00 & 0.83 & 1.19   & -- & -- & -- & -- \\
		Aug   & -- & --   & --   & --      & 109 & 1.20 & 1.06 & 1.37   & 1438 & 1.13 & 1.00 & 1.3  \\
		Sep   & 840 & 0.99 & 0.87 & 1.11   & 488 & 0.96 & 0.83 & 1.10   & 225 & 1.50 & 1.32 & 1.76 \\
		Oct   & 254 & 0.84 & 0.77 & 0.92   & 853 & 0.88 & 0.74 & 1.06   & -- & -- & -- & -- \\
		Nov   & 2596 & 0.92 & 0.74 & 1.25  & --  & --   & --   & --     & -- & --  & -- & --   \\
		Dec   & 3918 & 1.31 & 1.03 & 1.57  & 261 & 0.89 & 0.79 & 1.02   & --& --& -- & -- \\
		\bottomrule
	\end{tabular}
\end{table*}

\begin{table*}
	\centering
	\caption{Monthly median and quartile seeing values measured by the French DIMM at the North-2 Point (10m tower) from 2023 to 2024. A dash (–) denotes months with no available data.}
	\label{tab:2023-2024}
	\begin{tabular}{l *{4}{c}| *{4}{c}}
		\toprule
		Month & \multicolumn{4}{c|}{2023} & \multicolumn{4}{c}{2024} \\
		\cmidrule(lr){2-5} \cmidrule(lr){6-9}
		& Count & Median & 25\%  & 75\%  & Count & Median & 25\%  & 75\% \\
		\midrule
		Jan   & --    & --     & --    & --    & 69301 & 0.71   & 0.80  & 1.33  \\
		Feb   & --    & --     & --    & --    & 25970 & 0.72   & 0.73  & 0.98  \\
		Mar   & --    & --     & --    & --    & 24245 & 0.84   & 0.96  & 1.31  \\
		Apr   & --    & --     & --    & --    & 13409 & 0.86   & 0.77  & 1.10  \\
		May   & --    & --     & --    & --    & 18299 & 0.84   & 0.96  & 1.27  \\
		Jun   & --    & --     & --    & --    & 7290  & 0.97   & 0.82  & 0.97  \\
		Jul   & --    & --     & --    & --    & --    & --     & --    & --  \\
		Aug   & --    & --     & --    & --    & --    & --     & --    & --  \\
		Sep   & --    & --     & --    & --    & --    & --     & --    & --  \\
		Oct   & --    & --     & --    & --    & --    & --     & --    & --  \\
		Nov   & --    & --     & --    & --    & --    & --     & --    & --    \\
		Dec   & 59013  & 0.79   & 0.62  & 0.98 & --    & --     & --    & --  \\
		\bottomrule
	\end{tabular}
\end{table*}

Tables~\ref{tab:monthly_stats} and~\ref{tab:2023-2024} present the monthly statistics of median and quartile seeing values measured by the Nanjing DIMM at a height of 6\,m (North-1 Point) and the French DIMM at 10\,m (North-2 Point). A dash (--) indicates months for which no data were available.

\begin{figure}
	\includegraphics[width=\columnwidth]{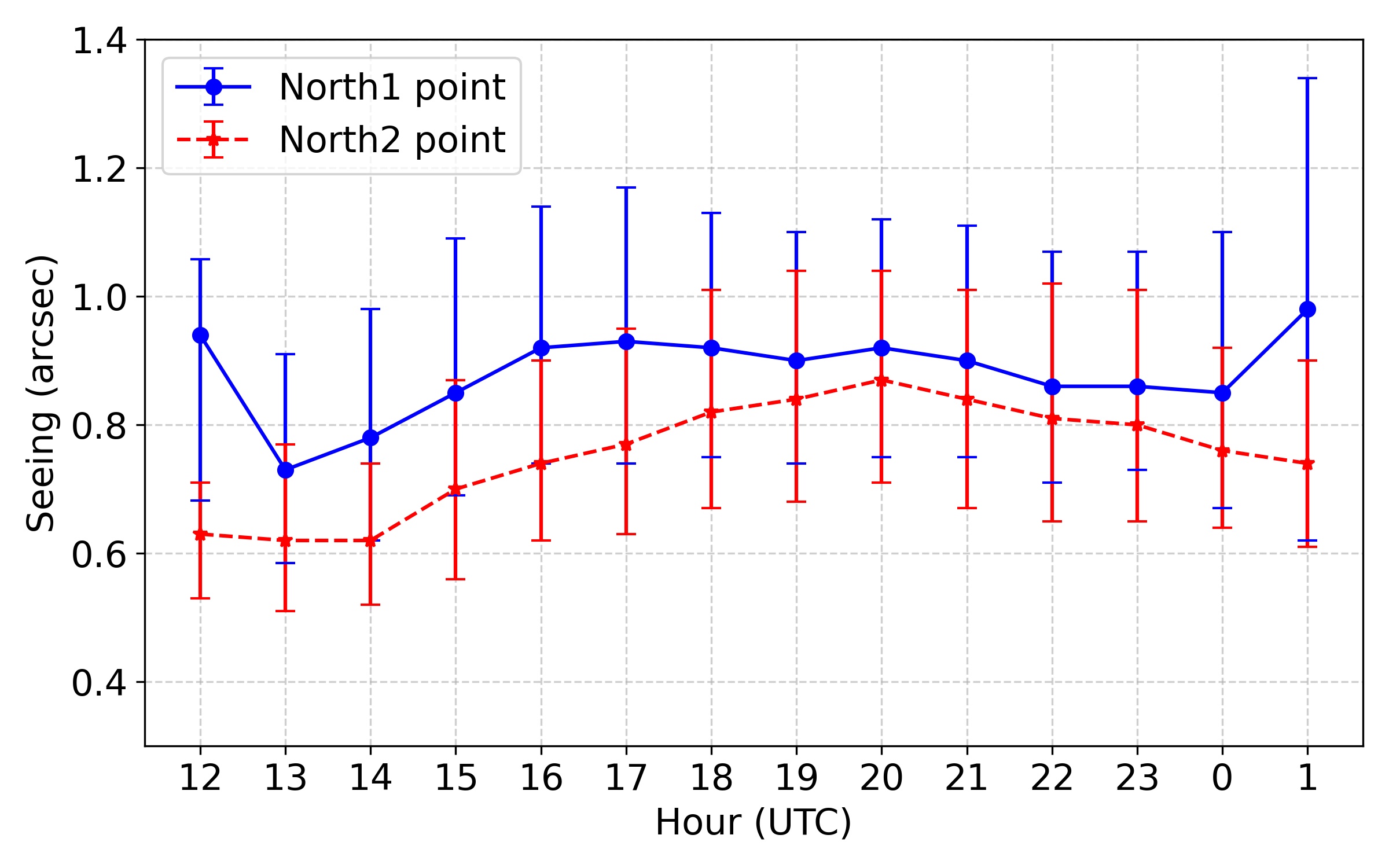}
	\caption{The hourly distribution of median seeing, with error bars representing the 50\% confidence interval.}
	\label{fig:seeing_hour}
\end{figure}

Figure~\ref{fig:seeing_hour} shows the hourly statistics of seeing at the North-1 and North-2 Points of the Muztagh-Ata site under cloud-free conditions. The results indicate that seeing conditions are generally more favorable during the early evening after sunset and again in the pre-dawn hours, while a noticeable degradation is observed around midnight. This pattern may be attributed to continued surface cooling after sunset, increasing wind speeds, and the development of vertical wind shear throughout the night.

\begin{figure*}
	\centering
	\includegraphics[width=\linewidth]{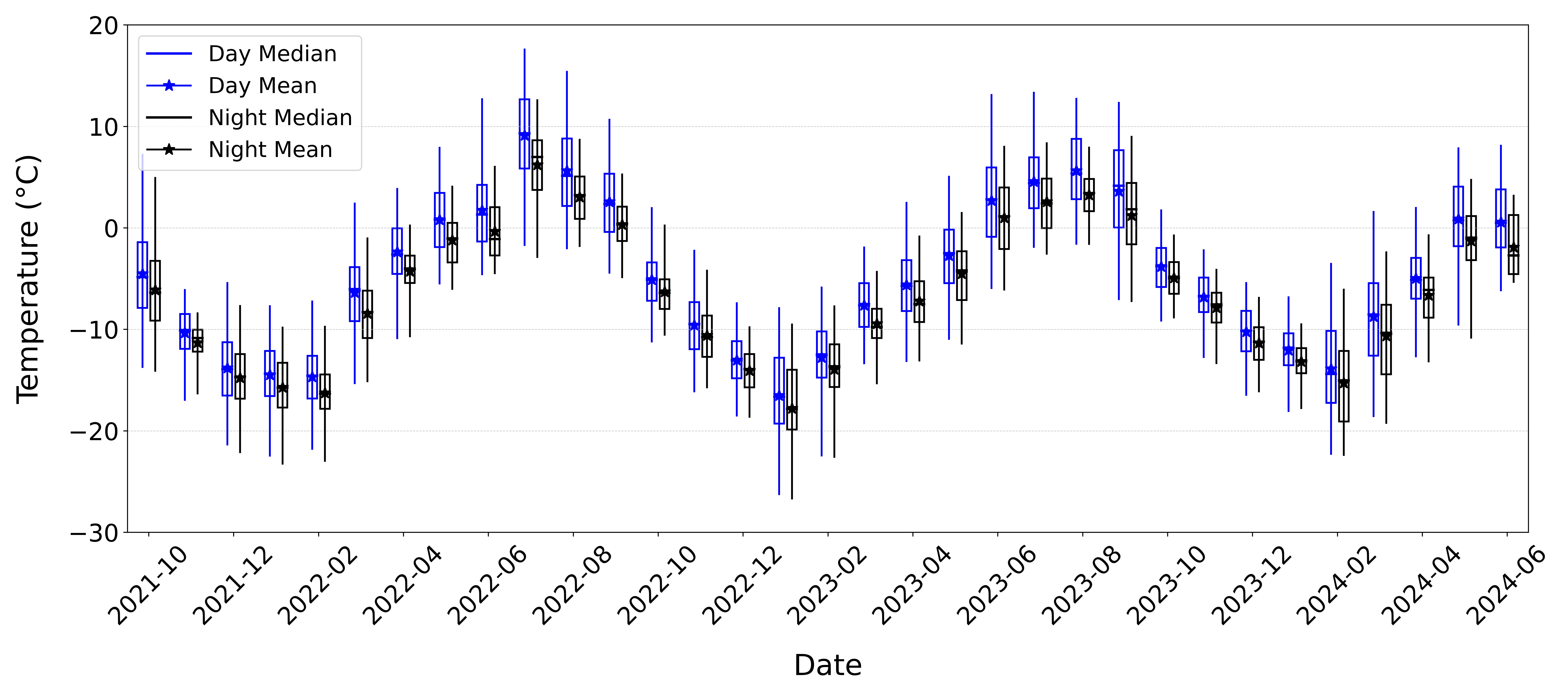}
	\caption{Monthly box‐and‐whisker plots of daytime and nighttime temperature distributions at the Muztagh‐Ata site from September 2021 to July 2024. The box spans the interquartile range (25th–75th percentiles) and the whiskers indicate the 1st–99th percentile range.}
	\label{fig:Temperature_DayNight}
\end{figure*}
\begin{figure}
	\includegraphics[width=\columnwidth]{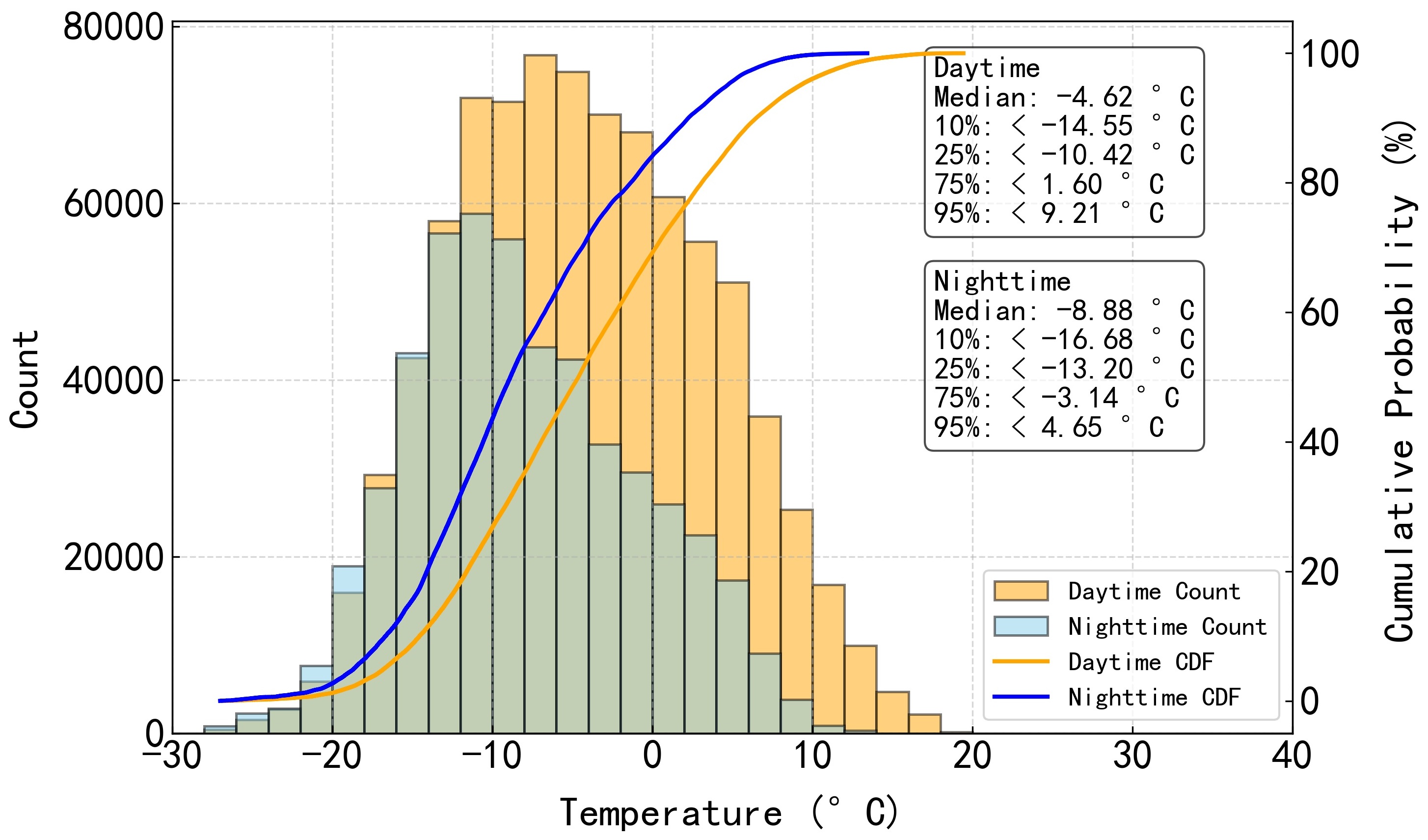}
	\caption{Statistical distributions and cumulative probability distributions of daytime and nighttime temperature at the Muztagh-Ata site.}
	\label{fig:Temperature_DayNight_Stats}
\end{figure}

\subsection{Meteorological Statistics and Air stability}
At the North-1 Point of the Muztagh-Ata site, a meteorological observation system was installed on a 30-meter tower. Data collection began in October 2021 and continued through June 2024, yielding a total of 963 valid observing days. After averaging the raw measurements at one-minute intervals, approximately 1,360,000 meteorological records were obtained.

At the North-2 Point, a PC-4A environmental monitoring unit was deployed as an automated weather station. Data collection began in January 2024, and by September 2024, a total of 180 valid observing days had been recorded, resulting in approximately 247,200 one-minute-resolution data points. It should be noted that no data were collected between 12 July and 10 September 2024.

\subsubsection{Temperature}

\begin{figure}
	\centering
	\includegraphics[width=\linewidth]{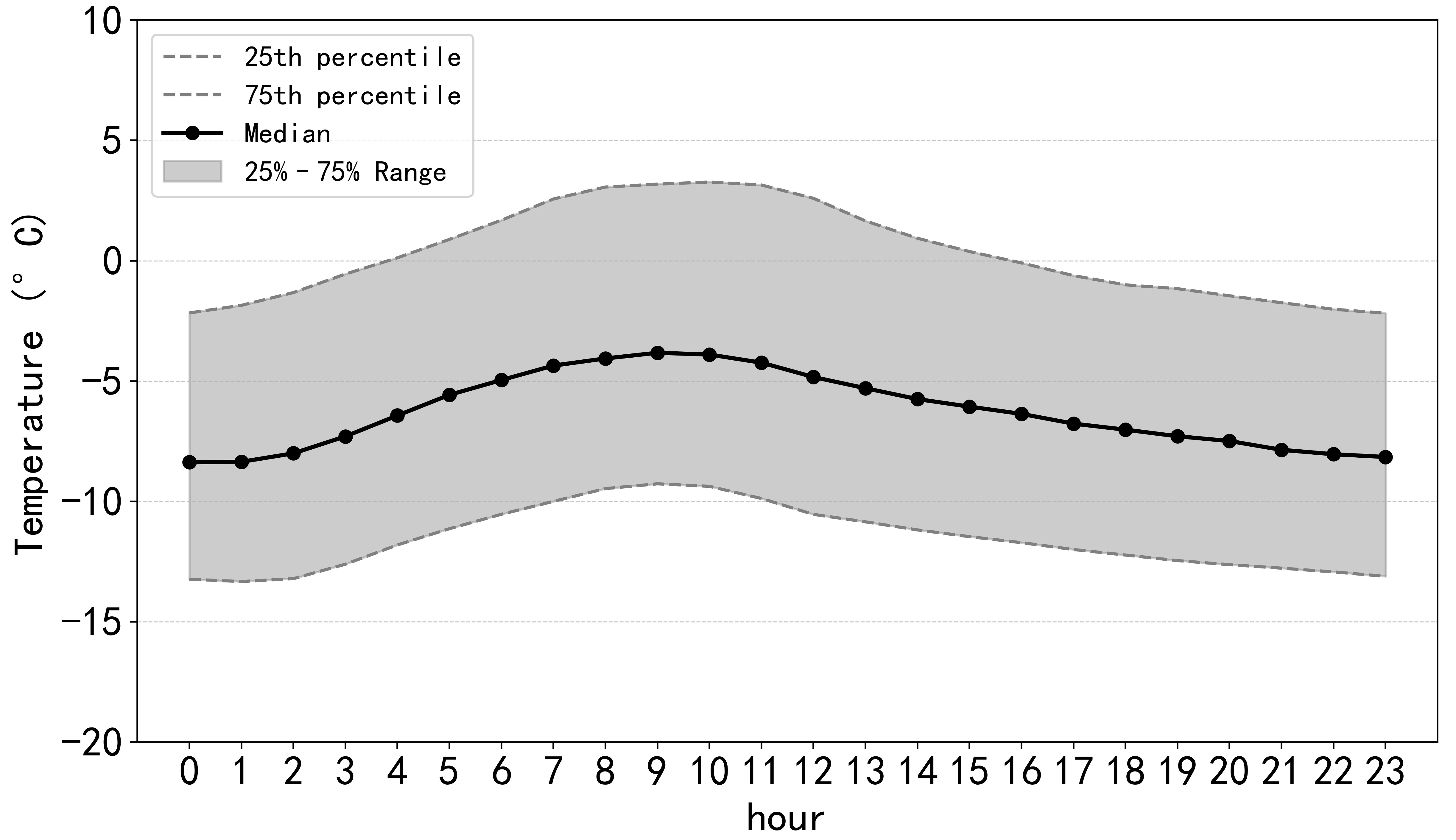}
	\caption{Diurnal variation of the median temperature at the Muztagh-Ata site. Error bars denote the 25th and 75th percentiles.}
	\label{fig:Temperaturehour}
\end{figure}

Figure~\ref{fig:Temperature_DayNight} presents monthly box-and-whisker plots of daytime and nighttime temperature distributions at the Muztagh-Ata site from September 2021 to June 2024. In each plot, the box represents the interquartile range (25th–75th percentiles), and the whiskers extend from the 1st to the 99th percentile. In all months, daytime temperatures are consistently higher than nighttime values, with a full range spanning from –27\,$^\circ$C to 18.5\,$^\circ$C. A clear seasonal trend is evident, with January being the coldest month and July the warmest.
Figure~\ref{fig:Temperature_DayNight_Stats} displays histograms and cumulative probability distributions of daytime and nighttime temperatures over the same period. The median daytime temperature is $-4.65\,^\circ$C, while the median nighttime temperature further decreases to $-8.72\,^\circ$C, indicating that nighttime surface cooling at the site is both significant and persistent.
Figure~\ref{fig:Temperaturehour} shows the diurnal variation of median temperature at the site. Error bars denote the 25th and 75th percentiles, illustrating the typical daily cycle. During daylight hours—particularly from 06:00 to 12:00—near-surface air temperature increases rapidly with solar radiation, reaching a distinct peak before gradually decreasing throughout the night.

\subsubsection{Air Stability}
Nighttime temperature variation is a critical parameter for evaluating overall seeing performance, as fluctuations in temperature over time and space directly affect the thermodynamic stability of the near-surface atmosphere. Larger nighttime temperature variation typically indicates weaker atmospheric stability, which in turn leads to degraded optical seeing~\citep{deng2021lenghu}.

\begin{figure}
	\includegraphics[width=1\columnwidth]{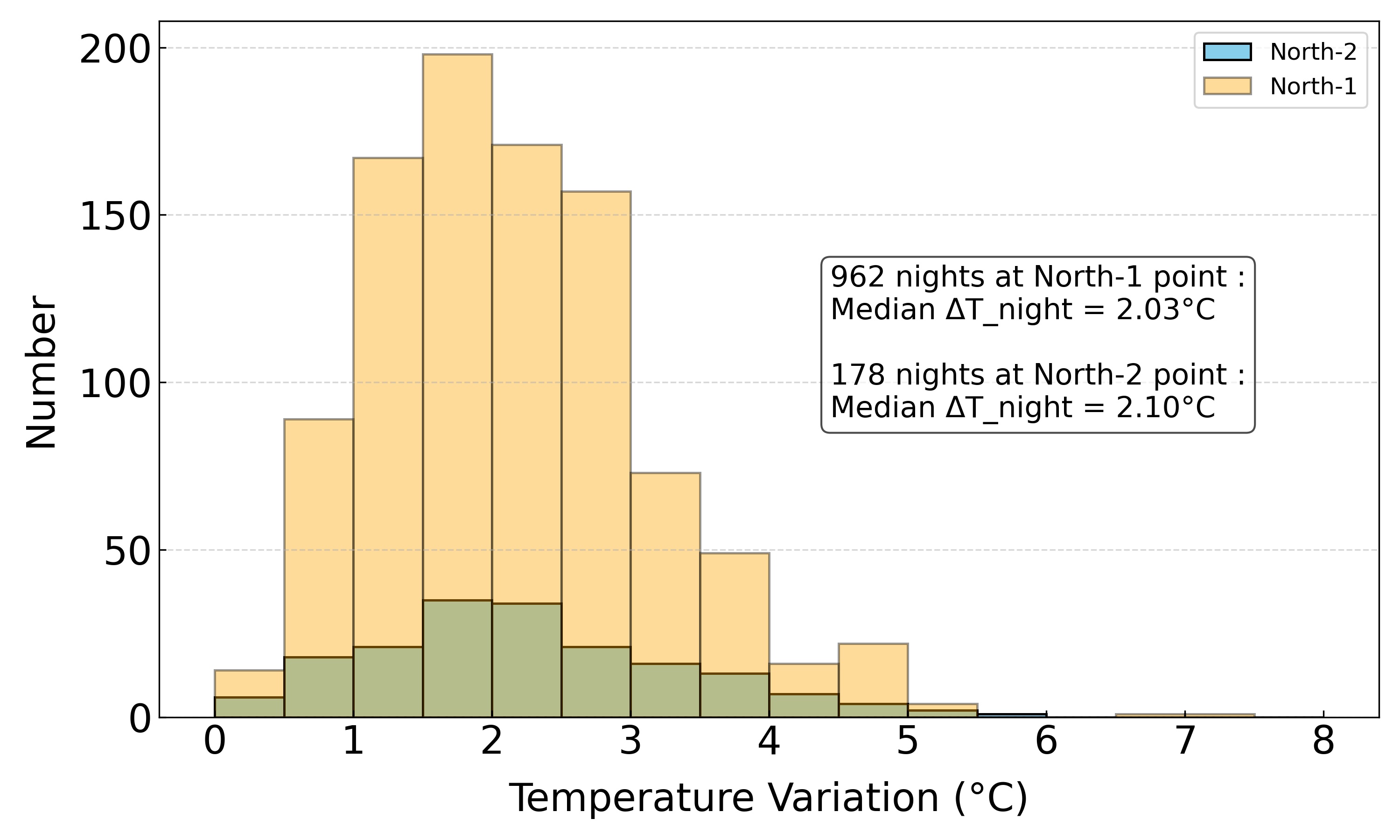}
	\caption{The upper panel shows the nighttime temperature variation range (10\%-90\%) statistics for the North-2 site, while the lower panel shows the nighttime temperature variation (10\%-90\%) range statistics for the North-1 site. "Total number of nights" represents the total number of nights during the data collection period.}
	\label{fig:fig9}
\end{figure}

Figure~\ref{fig:fig9} (top panel) shows the distribution of nighttime temperature variation at the North-2 Point of the Muztagh-Ata site. Over the observation period, the median diurnal temperature range reached $7.4\,^\circ\mathrm{C}$ (10th–90th percentile), while the median nighttime temperature variation was only $2.1\,^\circ\mathrm{C}$. The bottom panel presents data from the North-1 Point, where the median nighttime variation was slightly lower at $2.03\,^\circ\mathrm{C}$.

Compared with other world-class astronomical sites, this feature represents a significant advantage. For example, nighttime temperature variation reaches $6.8\,^\circ\mathrm{C}$ at Mauna Kea (Hawaii)~\citep{schock2009thirty}, $3.6\,^\circ\mathrm{C}$ at Cerro Paranal (Chile)~\citep{ESO1999}, and $2.7\,^\circ\mathrm{C}$ at Lenghu (Qinghai)~\citep{deng2021lenghu}. The remarkably small nighttime temperature variation observed at both North-1 and North-2 underscores the exceptional atmospheric stability of the Muztagh-Ata site—an essential condition for achieving high-quality ground-based optical observations.

\subsubsection{wind speed}
\begin{figure}
	\centering
	\includegraphics[width=\linewidth]{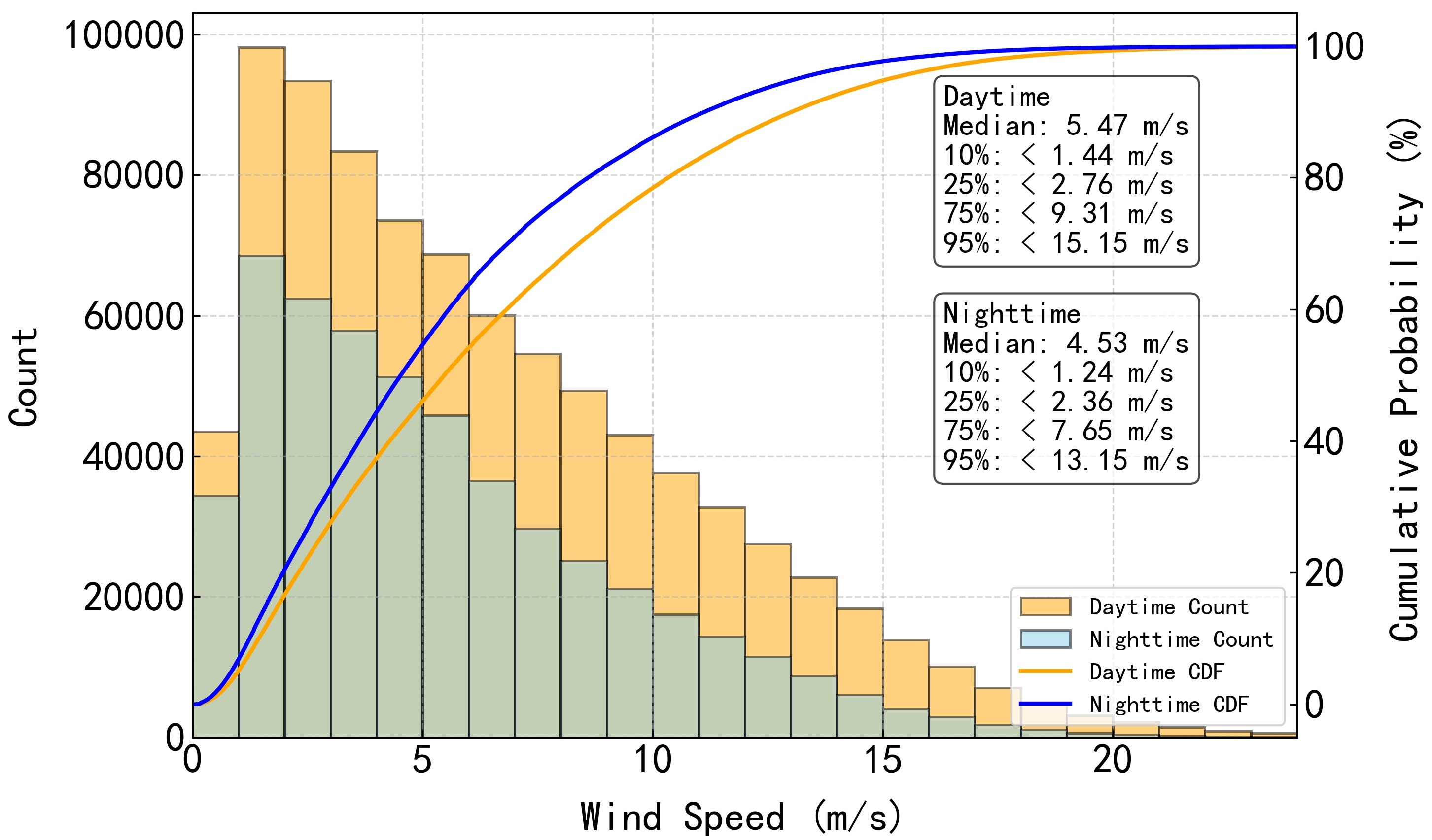}
	\caption{Statistical distributions and cumulative probability distributions of daytime and nighttime wind speeds at the Muztagh-Ata site.}
	\label{fig:windspeedfig5}
\end{figure}
\begin{figure}
	\centering
	\includegraphics[width=\linewidth]{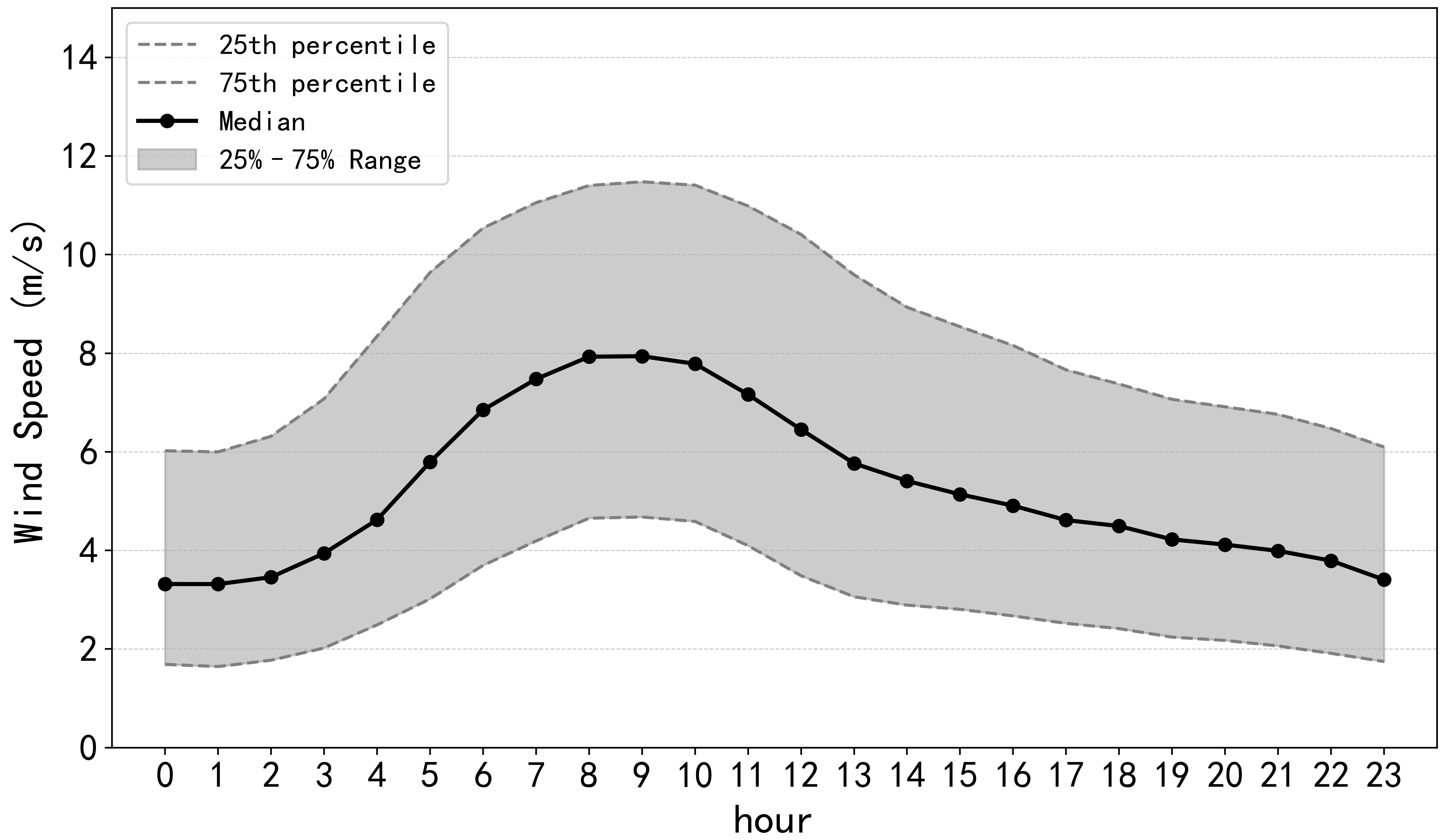}
	\caption{Hourly wind speed statistical distribution at the Muztagh-Ata site during the observation period. }
	\label{fig:windspeedfi}
\end{figure}
\begin{figure*}
	\centering
	\includegraphics[width=\linewidth]{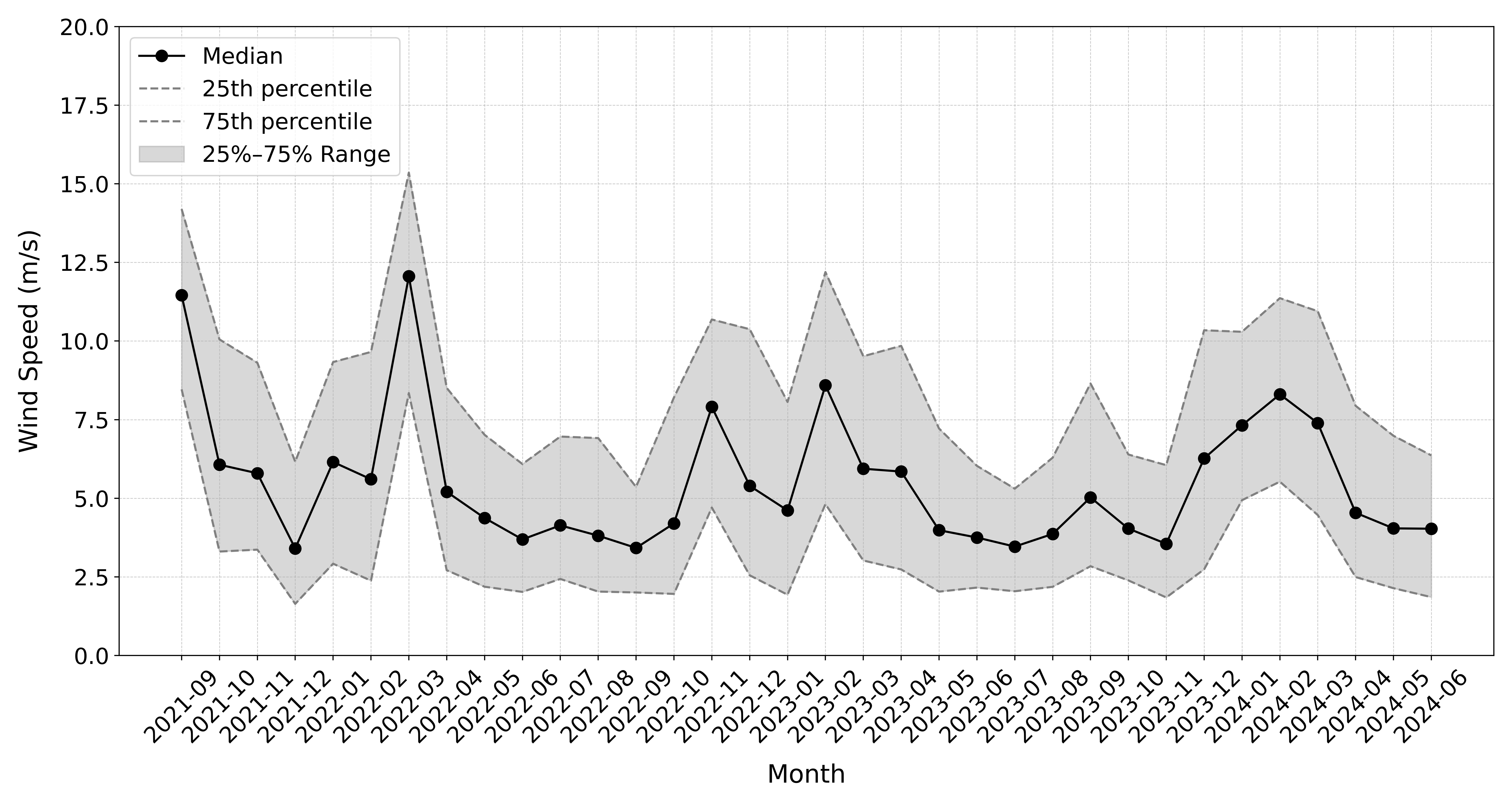}
	\caption{Monthly median wind speeds and 25th–75th percentile ranges observed at the Muztagh-Ata site during the monitoring period. }
	\label{fig:monthly_wind_median}
\end{figure*}

Based on wind speed and direction data collected at the Muztagh-Ata site from September 2021 to July 2024, we conducted a detailed analysis of local wind field characteristics. As illustrated in Figure~\ref{fig:windspeedfig5}, the statistical and cumulative probability distributions of wind speed show clear diurnal variation. The median wind speed during daytime reaches $5.75\,\mathrm{m/s}$, while the nighttime value is $4.79\,\mathrm{m/s}$, indicating that wind speeds are generally stronger during the day than at night.

Figure~\ref{fig:windspeedfi} shows the statistical distribution of hourly wind speeds at the Muztagh-Ata site over the entire observation period. The results indicate that elevated wind speeds primarily occur from the afternoon to evening hours. This pattern is closely associated with strong solar heating of the plateau surface during daytime, which enhances vertical atmospheric instability. The resulting increase in the temperature gradient between the near-surface and the overlying atmosphere intensifies boundary-layer mixing, thereby giving rise to a characteristic diurnal wind structure.

Figure~\ref{fig:monthly_wind_median} presents the monthly median wind speeds and their interquartile ranges (25th–75th percentiles) from October 2021 to June 2024. Overall, wind speeds increase markedly from winter through early spring (December to March), decrease significantly in summer (June to August), and gradually recover in autumn, accompanied by relatively low variability.

\begin{figure}
	\centering
	\includegraphics[width=1.05\linewidth]{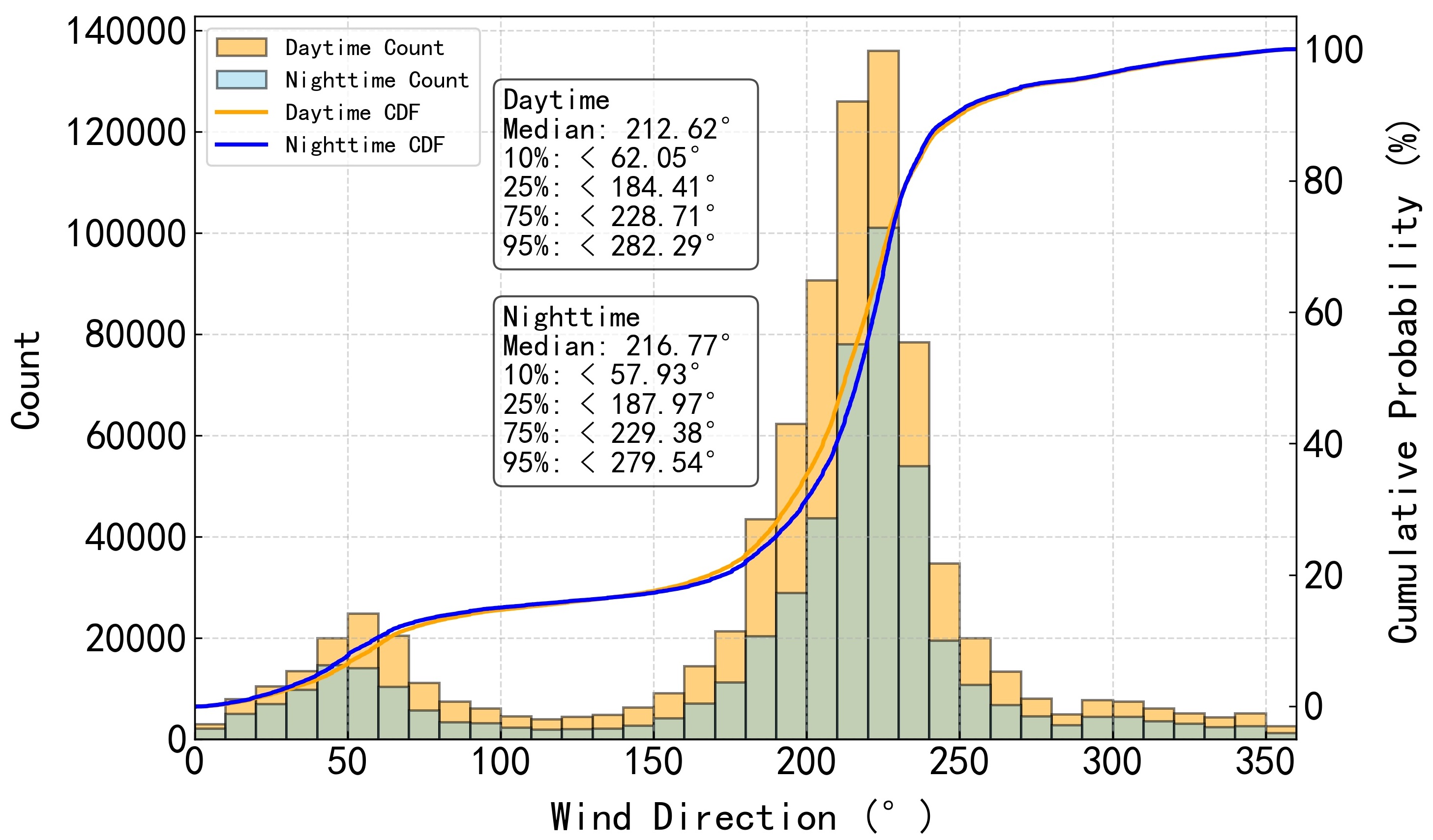}
	\caption{Statistical and cumulative distributions of daytime and nighttime wind directions at the Muztagh-Ata site. }
	\label{fig:WindDirection_DayNight_Stats}
\end{figure}
\begin{figure}
	\centering
	\includegraphics[width=1.05\linewidth]{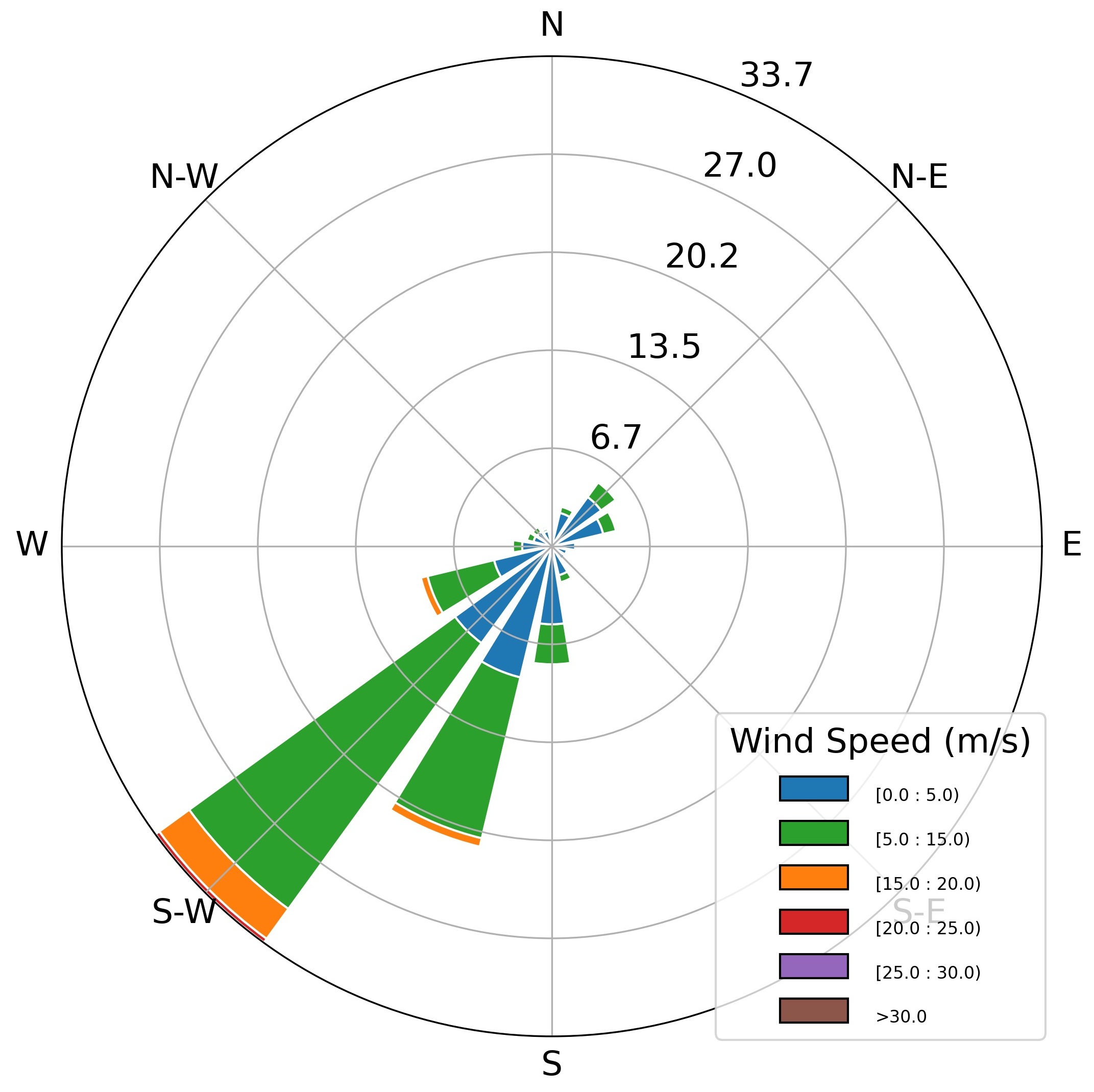}
	\caption{windrose diagrams at the Muztagh-Ata site.}
	\label{fig:windrose}
\end{figure}

Figure~\ref{fig:WindDirection_DayNight_Stats} and Figure~\ref{fig:windrose} present the wind rose and the distributions of wind direction during daytime and nighttime at the Muztagh-Ata site, respectively. Although some directional dispersion is evident, the prevailing wind directions are consistently concentrated within the southwest quadrant, ranging from $210^\circ$ to $300^\circ$. Approximately 72\% of all measurements fall within this range, indicating a dominant southwest-to-northeast airflow. This pattern aligns with expectations given the site's surrounding topography and the channeling effects imposed by nearby mountain terrain.

When considering the relatively lower nighttime wind speeds in conjunction with the stability of wind direction, the site exhibits notable aerodynamic stability during nighttime hours. Under such conditions—characterized by sustained wind direction and minimal speed fluctuations—the generation of local turbulence is effectively suppressed. This contributes positively to maintaining favorable seeing and stable observational conditions. The overall stability of the nighttime wind field is thus critical for achieving high-quality ground-based astronomical observations.

\subsubsection{Relative Humidity and Dew Point Temperature}

\begin{figure}
	\centering
	\includegraphics[width=\linewidth]{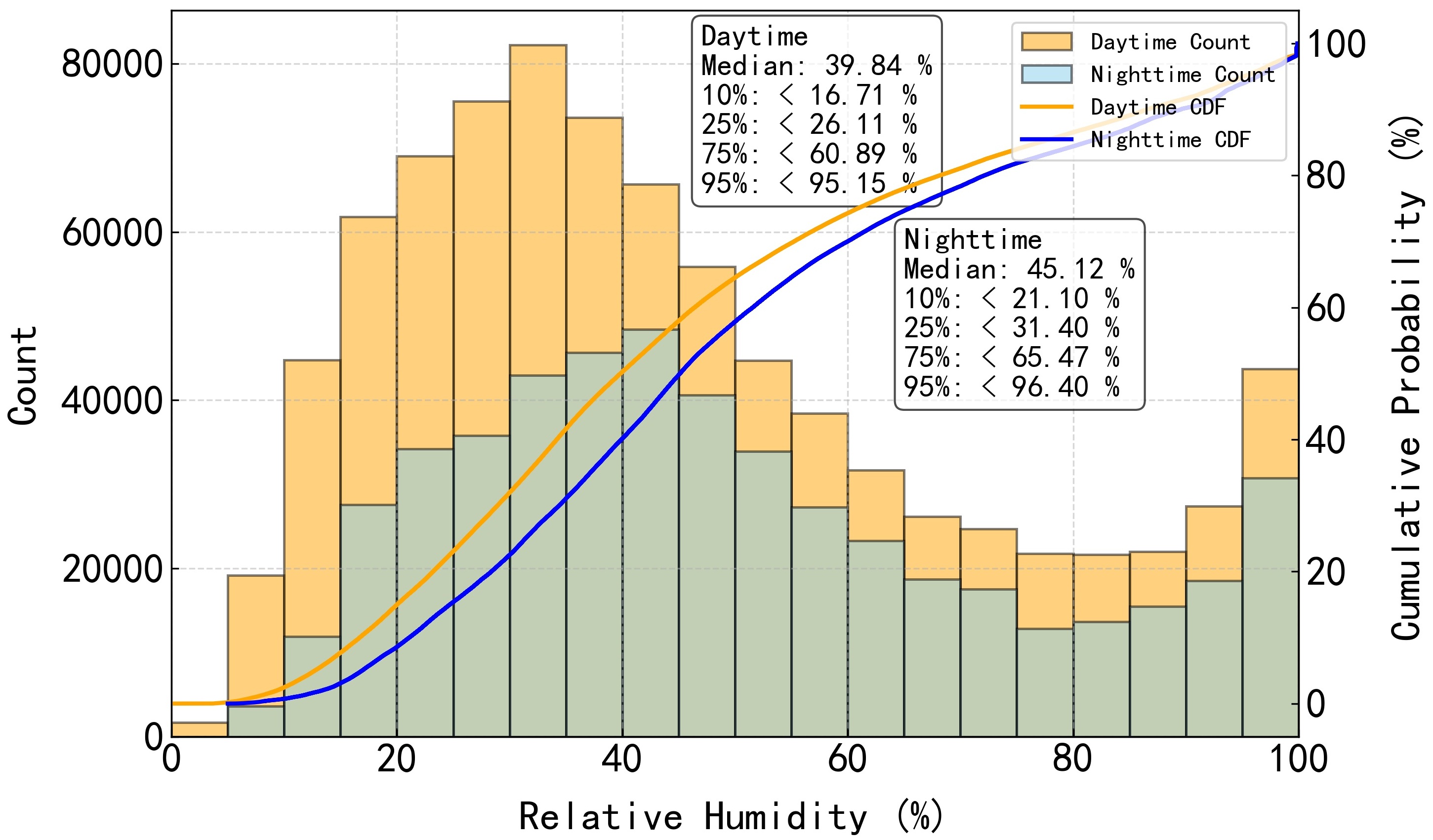}
	\caption{Histogram and cumulative distribution functions of daytime and nighttime relative humidity at the Muztagh-Ata site.}
	\label{fig:Humidity_DayNight_Stats}
\end{figure}

As shown in Figure~\ref{fig:Humidity_DayNight_Stats}, the diurnal variation of relative humidity (RH) at the Muztagh-Ata site from September 2021 to June 2024 indicates that the median nighttime RH ($45.12\%$) is significantly higher than the daytime value ($39.84\%$). This difference is mainly caused by lower nighttime temperatures, which reduce the temperature–dew point spread. Even with stable moisture content, this leads to higher RH.

\begin{figure}
	\centering
	\includegraphics[width=\linewidth]{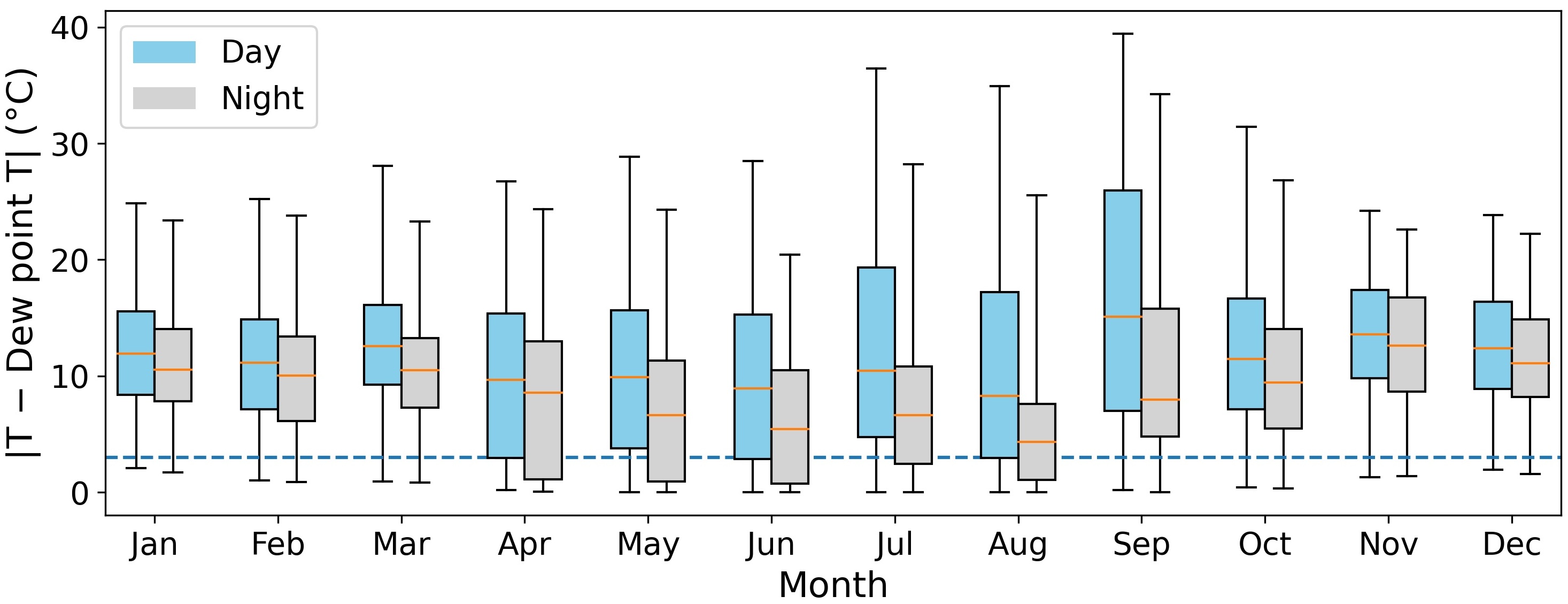}
	\caption{Monthly distribution of the absolute difference between temperature and dew point during daytime and nighttime at the Muztagh-Ata site. The dashed line marks the 3\,°C warning threshold.}
	\label{fig:temp_dewpoint_monthly}
\end{figure}

Figure~\ref{fig:temp_dewpoint_monthly} supports this interpretation, showing that the nighttime temperature–dew point spread is consistently smaller than during the day throughout the year. A threshold of $3^{\circ}\mathrm{C}$ is marked as a dew/frost risk line. When the spread drops below this value, the risk of condensation or frost on instruments increases. From April to August, nighttime conditions frequently approach or fall below this threshold, indicating elevated frosting risk during these months.

\subsubsection{air pressure}

\begin{figure}
	\centering
	\includegraphics[width=\linewidth]{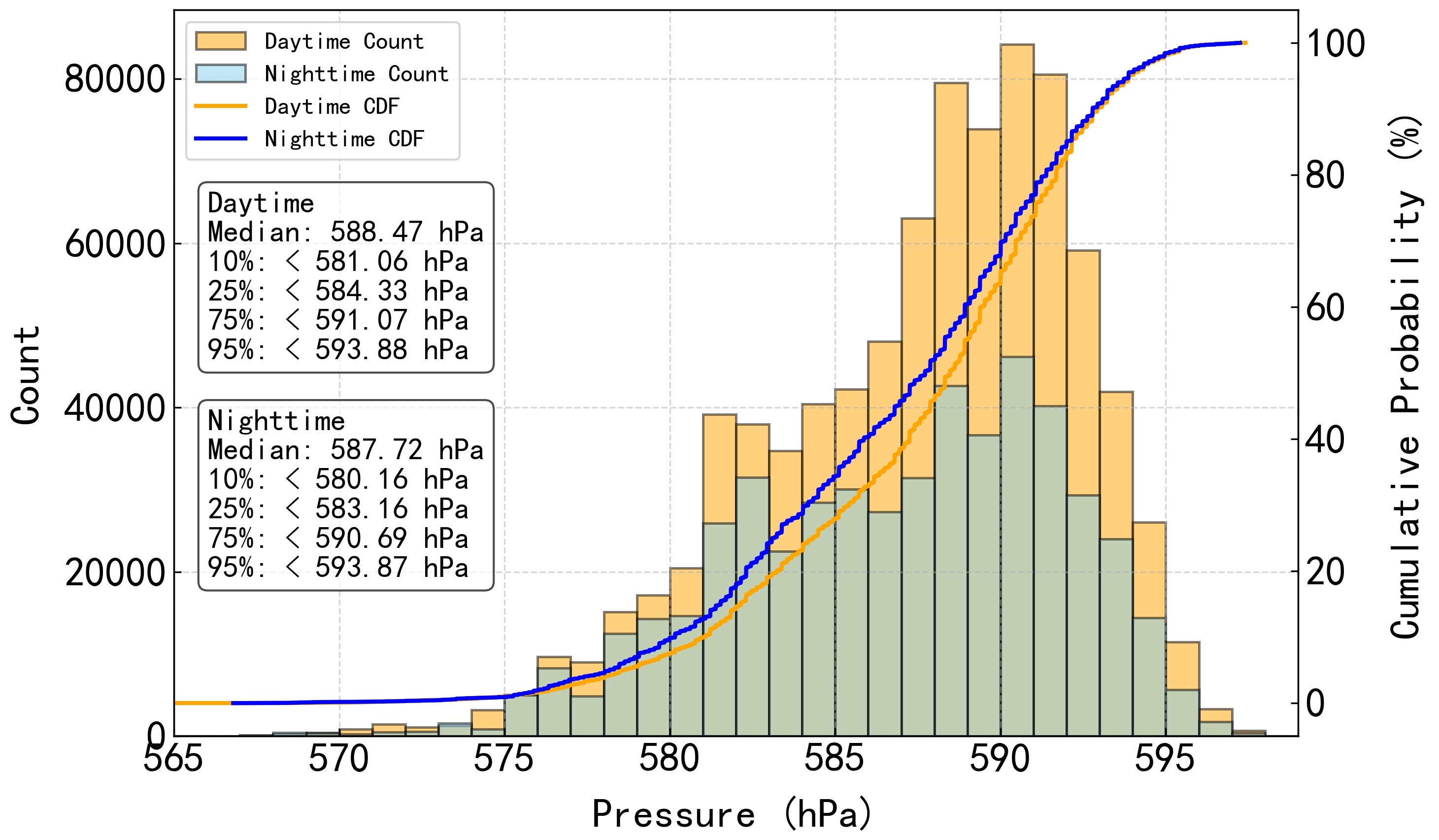}
	\caption{Histogram and cumulative distribution functions of daytime and nighttime surface air pressure at the Muztagh-Ata site.}
	\label{fig:Pressure_DayNight_Stats}
\end{figure}

Figure~\ref{fig:Pressure_DayNight_Stats} shows the statistical and cumulative probability distributions of surface air pressure during daytime and nighttime at the Muztagh-Ata site from October 2021 to June 2024. The results reveal a high degree of similarity between daytime and nighttime pressure distributions, with slightly higher median pressure during the day (588.47\,hPa) compared to night (587.72\,hPa). Overall, surface pressure values are predominantly concentrated in the range of 580--595\,hPa, exhibiting minimal variability. This indicates a highly stable pressure regime at the high-altitude Muztagh-Ata site, which is conducive to sustained high-precision astronomical observations.

\subsection{Nighttime Cloud Cover and Observing Time}
Nighttime sky clarity is a critical factor in selecting sites for optical and infrared observatories, as cloud cover directly determines the available time for scientific observation. To evaluate the observational potential of the Muztagh-ata site, we analyzed nighttime cloud cover data collected from January 1, 2017, through June 2024. The data were acquired using a ground-based all-sky camera (ASC) equipped with a Canon DSLR, which captures full-sky images (3456x5184 pixels) every 2 to 5 minutes.Following the classification scheme of \citep{2008SPIE.7012E..24S}, each image is categorized based on cloud distribution within defined zenith distance circles. In this classification scheme, the region with zenith angles from $0^\circ$ to $44.7^\circ$ is defined as the inner circle, and the region from $44.7^\circ$ to $65^\circ$ as the outer circle.
\begin{itemize}
	\item \textbf{Clear:} No cloud.
	\item \textbf{Outer:} No cloud within the inner circle, and cloud within the outer circle.
	\item \textbf{Inner:} No more than 50\% cloud within both the inner and outer circles.
	\item \textbf{Covered:} Cloud coverage within inner + outer circle exceeds 50\%.
\end{itemize}

This paper extends the previous work of \citep{2023RAA....23d5015X}, which analyzed data from 2017 to 2021. Table~\ref{tab:yearly_cloud_stats} presents the yearly percentage distribution of sky conditions and the corresponding data acquisition rates at the Muztagh-ata site.
We define Available Observing Time (AOT) as the sum of "Clear" and "Outer" conditions, and Unavailable Observing Time (UOT) as the sum of "Inner" and "Covered" conditions. Across the entire monitoring period, the AOT averaged approximately 64\%, while the UOT was about 36\%. The overall data acquisition rate averaged 87.3\%.

\begin{figure}
	\centering
	\includegraphics[width=\linewidth]{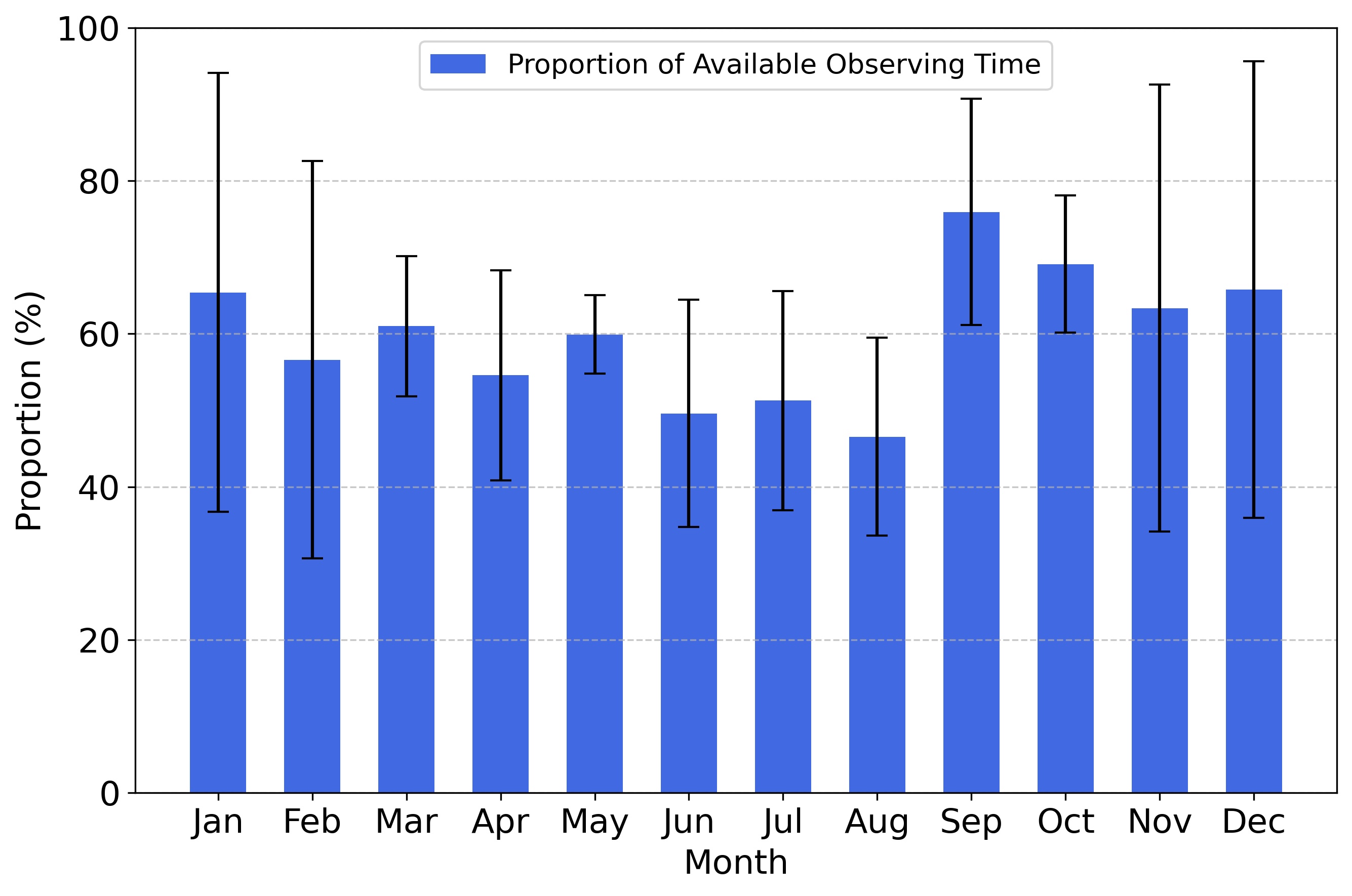}
	\caption{Monthly average of Available Observing Time (AOT) at the Muztagh-ata site from January 2017 to June 2023. The AOT is the sum of 'Clear' and 'Outer' sky conditions. The error bars represent the standard deviation for each month over the observation period.}
	\label{fig:monthly_sky_proportion}
\end{figure}

\begin{table}
	\centering
	\caption{Statistics of nighttime all-sky image classification and data acquisition at the Muztagh-Ata site (2017–2024), with data from 2017–2021 summarized in~\citep{2023RAA....23d5015X}.}
	\label{tab:yearly_cloud_stats}
	\begin{tabular}{lcccccc}
		\hline
		Year & Clear & Outer & Inner & Covered & Data Acquisition \\
		\hline
		2017 & 55\% & 6\% & 7\% & 32\% & 92.7\% \\
		2018 & 56\% & 10\% & 5\% & 29\% & 98.2\% \\
		2019 & 57\% & 7\% & 6\% & 30\% & 81.1\% \\
		2020 & 60\% & 5\% & 6\% & 29\% & 84.3\% \\
		2021 & 64\% & 8\% & 2\% & 26\% & 92.4\% \\
		2022 & 54\% & 9\% & 7\% & 30\% & 74.8\% \\
		2023 & 56\% & 8\% & 7\% & 29\% & 80.1\% \\
		2024 & 55\% & 2\% & 7\% & 36\% & 95.0\% \\
		\hline
		Total & 57.1\% & 6.9\% & 5.8\% & 30.2\% & 87.3\% \\
		\hline
	\end{tabular}
\end{table}
To investigate the seasonal variation of cloud cover, we analyzed the monthly distribution of Available Observing Time (AOT) from January 2017 to June 2023. Figure~\ref{fig:monthly_sky_proportion} shows the monthly mean AOT values with standard deviations. The results reveal significant seasonal variability at the Muztagh-Ata site, with the highest AOT observed in autumn and winter (average $>60\%$) and relatively low standard deviations. In contrast, cloud cover increases notably in summer, especially in July and August.

To further assess the observational potential of the Muztagh-Ata site, we compared its  Astronomically Observable Time fraction with those of other established sites, including  Daocheng in Sichuan, Cerro Tolar, San Pedro Mártir and others. Table~\ref{tab:aot_comparison} summarizes the AOT fractions based on a unified classification scheme. The "AOT Criterion" column specifies the standards used to define astronomically usable time.~\citep{2008SPIE.7012E..24S, 2020RAA....20...85S, 2023RAA....23d5015X}.

\begin{table}
	\centering
	\caption{Comparison of available observing time fractions at different astronomical sites. All values are based on the “Clear + Outer” classification unless otherwise noted.}
	\label{tab:aot_comparison}
	\begin{tabular}{lcc}
		\hline
		Site & AOT (\%) & AOT Criterion \\
		\hline
		Muztagh-ata & 64.0 & Clear + Outer \\
		Daocheng & 65.0 & Clear + Outer + Inner \\
		Cerro Tolar & 83.7 & Clear + Outer \\
		Cerro Armazones & 86.8 & Clear + Outer \\
		Cerro Tolonchar & 77.8 & Clear + Outer \\
		San Pedro Mártir & 77.9 & Clear + Outer \\

		\hline
	\end{tabular}
\end{table}

\subsection{Correlation Between Seeing and Wind Parameters}

\begin{figure}
	\centering
	\includegraphics[width=\linewidth]{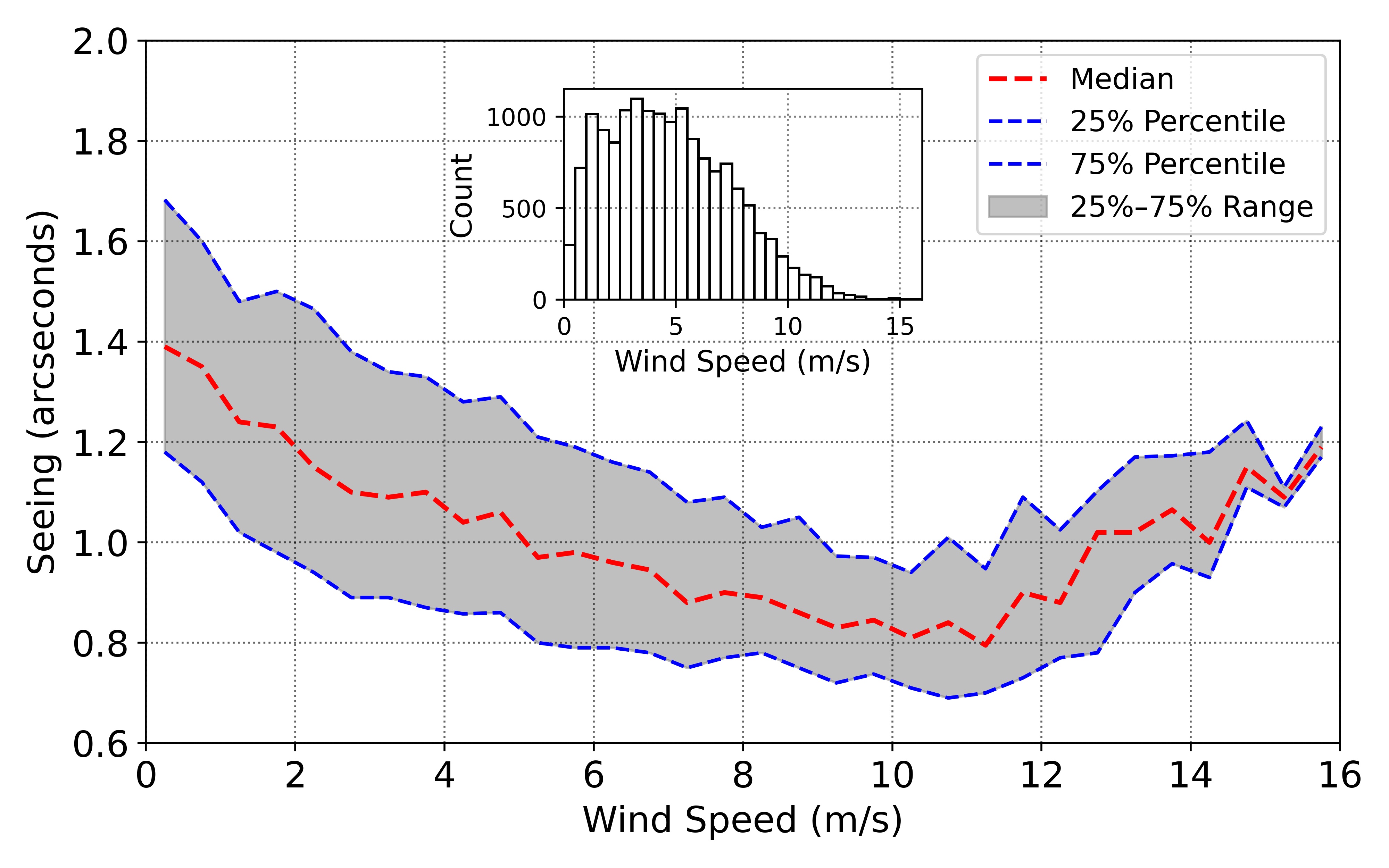}
	\caption{Relationship between seeing and average average wind speed at the North-1 point. The red dashed line indicates the median seeing, and the gray shaded area represents the interquartile range (25\%--75\%).}
	\label{fig:seeing_wind_north1}
\end{figure}

\begin{figure}
	\centering
	\includegraphics[width=\linewidth]{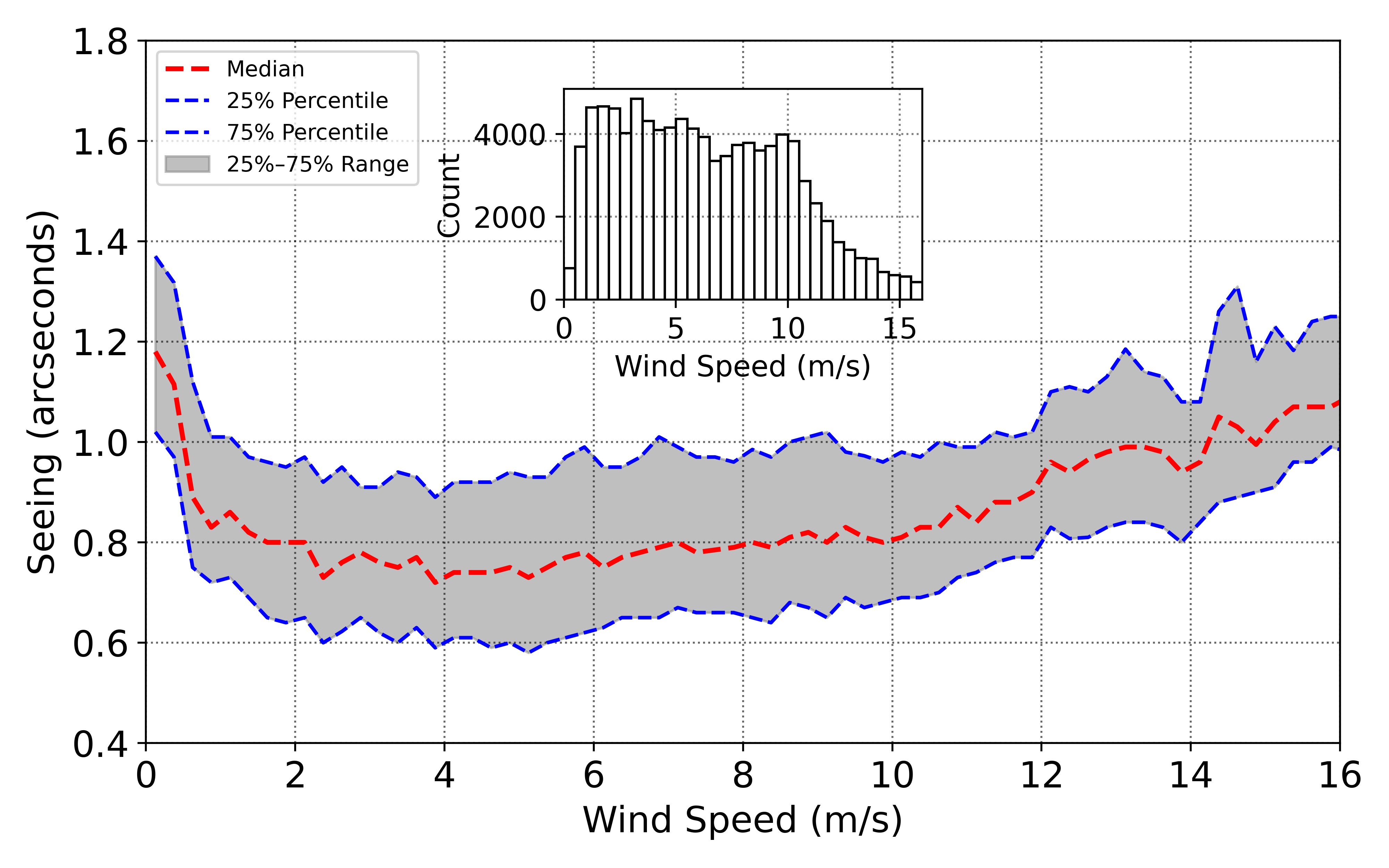}
	\caption{Relationship between seeing and average average wind speed at the North-2 point. The red dashed line indicates the median seeing, and the gray shaded area represents the interquartile range (25\%--75\%).}
	\label{fig:seeing_wind_north2}
\end{figure}

To further investigate the impact of wind speed on seeing, we analyzed the correlation between average wind speed and seeing at the North-1 and North-2 points. Figures~\ref{fig:seeing_wind_north1} and~\ref{fig:seeing_wind_north2} illustrate the variation of seeing with wind speed during the monitoring periods at North-1 and North-2, respectively. Each figure shows the median seeing (red dashed line), the interquartile range (25\%--75\%, shaded in gray), and the 25\% and 75\% percentiles (blue dashed lines). A pronounced U-shaped trend is evident at North-1: seeing improves with increasing wind speed in the range of 4--10\,m/s, but deteriorates rapidly beyond this range. A similar pattern is observed at North-2: seeing steadily improves with wind speeds from 0 to 4\,m/s, remains relatively stable between 4 and 10\,m/s, and then worsens significantly at higher speeds. These differences may arise from variations in sensor height, local topography, and observation periods, indicating that the relationship between wind and optical turbulence is both site- and time-dependent. Moderate wind speeds help suppress turbulence and stabilize seeing, while extremely low or high wind speeds tend to enhance atmospheric disturbances, thereby degrading observational quality.

\begin{figure}
	\centering
	\includegraphics[width=\linewidth]{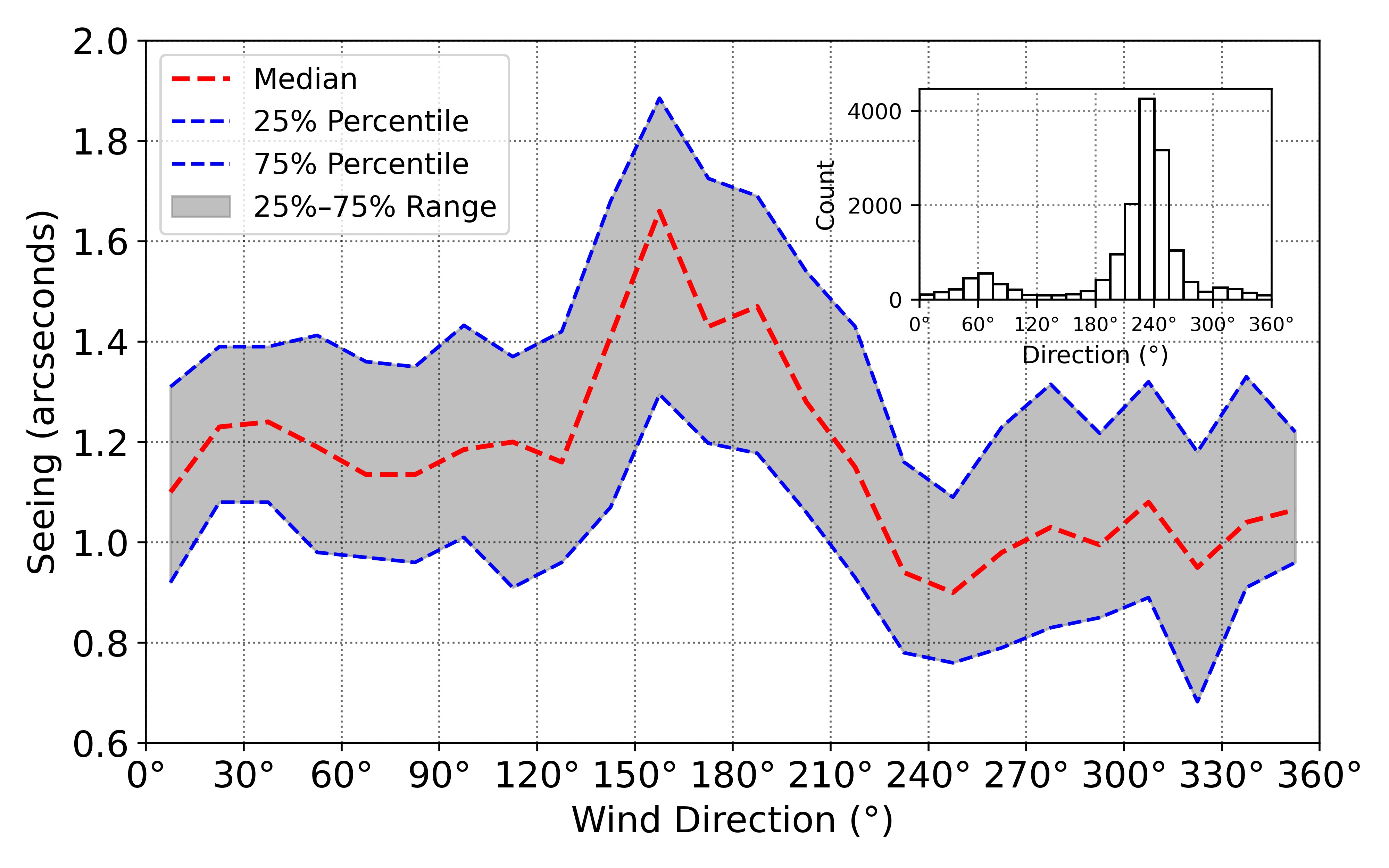}
	\caption{Relationship between seeing and average wind direction at the North-1 point. The red dashed line indicates the median seeing, and the gray shaded area represents the interquartile range (25\%--75\%).}
	\label{fig:wind_dir_vs_seeing}
\end{figure}

\begin{figure}
	\centering
	\includegraphics[width=\linewidth]{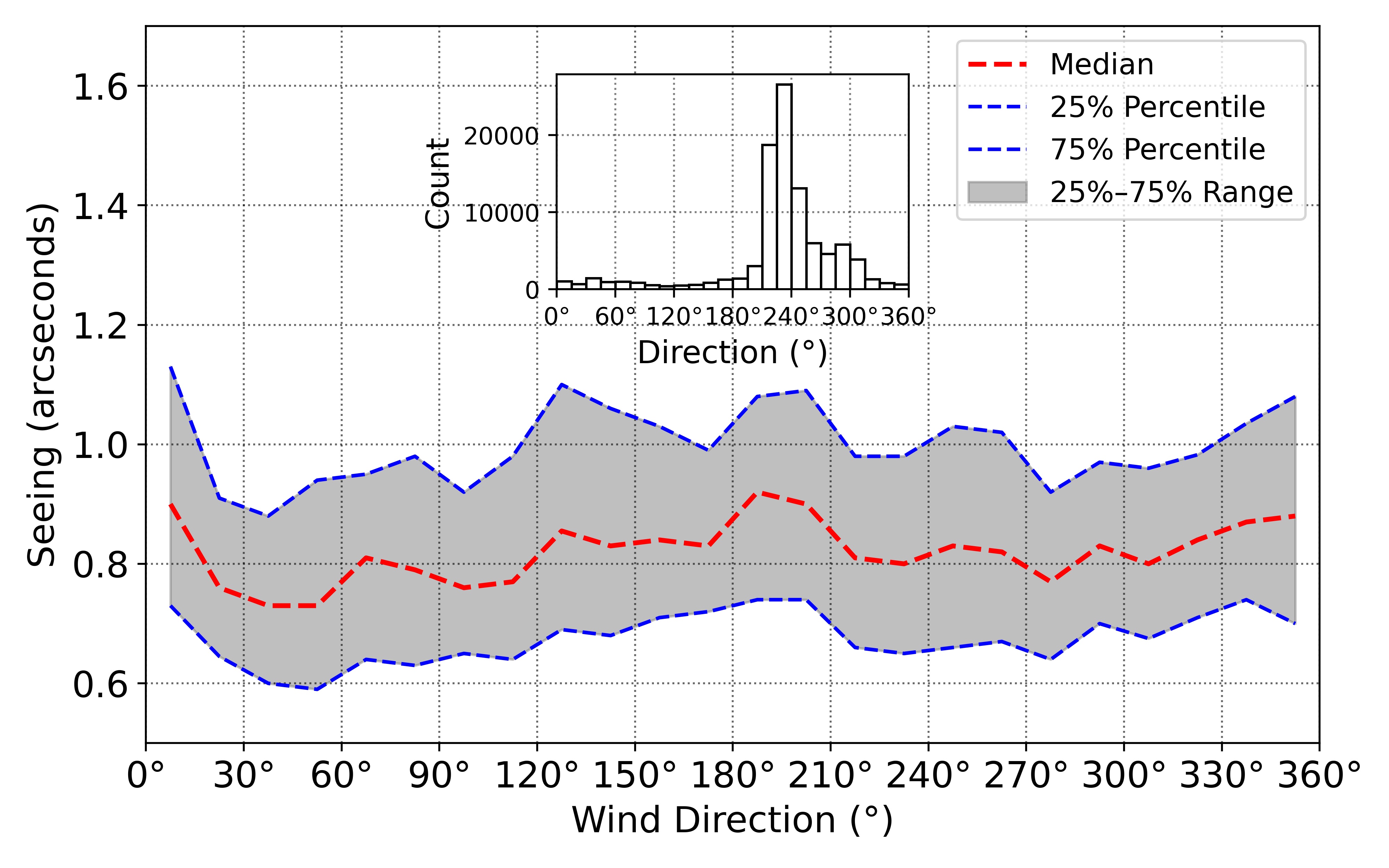}
	\caption{Relationship between seeing and average wind direction at the North-2 point. The red dashed line indicates the median seeing, and the gray shaded area represents the interquartile range (25\%--75\%).}
	\label{fig:North2wind_dir_vs_seeing}
\end{figure}

Figures~\ref{fig:wind_dir_vs_seeing} and~\ref{fig:North2wind_dir_vs_seeing} illustrate the relationship between seeing and wind direction at the North-1 and North-2 points of the Muztagh-Ata site. In both cases, the prevailing wind sector lies between 180$^\circ$ and 270$^\circ$ (southwest quadrant), where the median seeing remains low and the 25\%--75\% percentile range is narrow, suggesting that airflow from this direction interacts with the local topography to generate relatively weak turbulence. In contrast, although winds from 100$^\circ$--150$^\circ$ (southeast quadrant) occur less frequently, they are associated with noticeably higher median seeing and a broader percentile spread, indicating that flow through nearby mountain valleys likely introduces stronger shear and turbulence. Notably, the North-1 site shows a more sensitive response in seeing to changes in wind direction compared to the North-2 site, which exhibits a smoother trend—possibly due to the shorter observation period and incomplete seasonal coverage at North-2.

Vertical wind shear and temporal wind variation are key factors influencing astronomical seeing conditions \citep{zhi2024impact}. Wind shear generally describes the spatial gradient of wind speed or direction, with vertical wind shear referring specifically to changes in wind characteristics along the vertical axis. In contrast, temporal wind variation captures how wind speed and direction fluctuate over time.

In this study, we focus on vertical wind shear as an indicator of the rate of wind speed change with altitude.
The ERA-5 dataset, produced by the European Centre for Medium-Range Weather Forecasts (ECMWF), offers fifth-generation global atmospheric reanalysis data from 1940 to the present. It provides hourly atmospheric, oceanic, and land-surface variables at a spatial resolution of 0.25$^\circ$ $\times$ 0.25$^\circ$, making it well-suited for characterizing atmospheric dynamics at high resolution.
Following the approach described in \citep{Li2007,2024RAA....24a5006W}, we processed the ERA5 reanalysis data for the Muztagh-Ata site by converting geopotential heights at various levels into geometric heights. This allowed us to construct vertical atmospheric profiles of temperature, pressure, and wind speed.

By computing the differences in wind speed components between adjacent altitude levels and applying the Euclidean distance formula to obtain the total wind speed variation, followed by normalization using the vertical height difference, the vertical wind shear can be expressed as:
\begin{equation}
	S = \frac{\sqrt{(\Delta U)^2 + (\Delta V)^2}}{\Delta z}
\end{equation}
where \( S \) denotes the vertical wind shear, \( \Delta U \) and \( \Delta V \) represent the changes in wind speed components in the east–west and north–south directions, respectively, and \( \Delta z \) is the vertical separation between adjacent layers. The absolute value of shear is used to ensure consistency in magnitude interpretation. 
Larger values of vertical wind shear reflect more pronounced changes in wind velocity with altitude, which can intensify atmospheric turbulence and negatively affect seeing conditions.

\begin{figure}
	\centering
	\includegraphics[width=\linewidth]{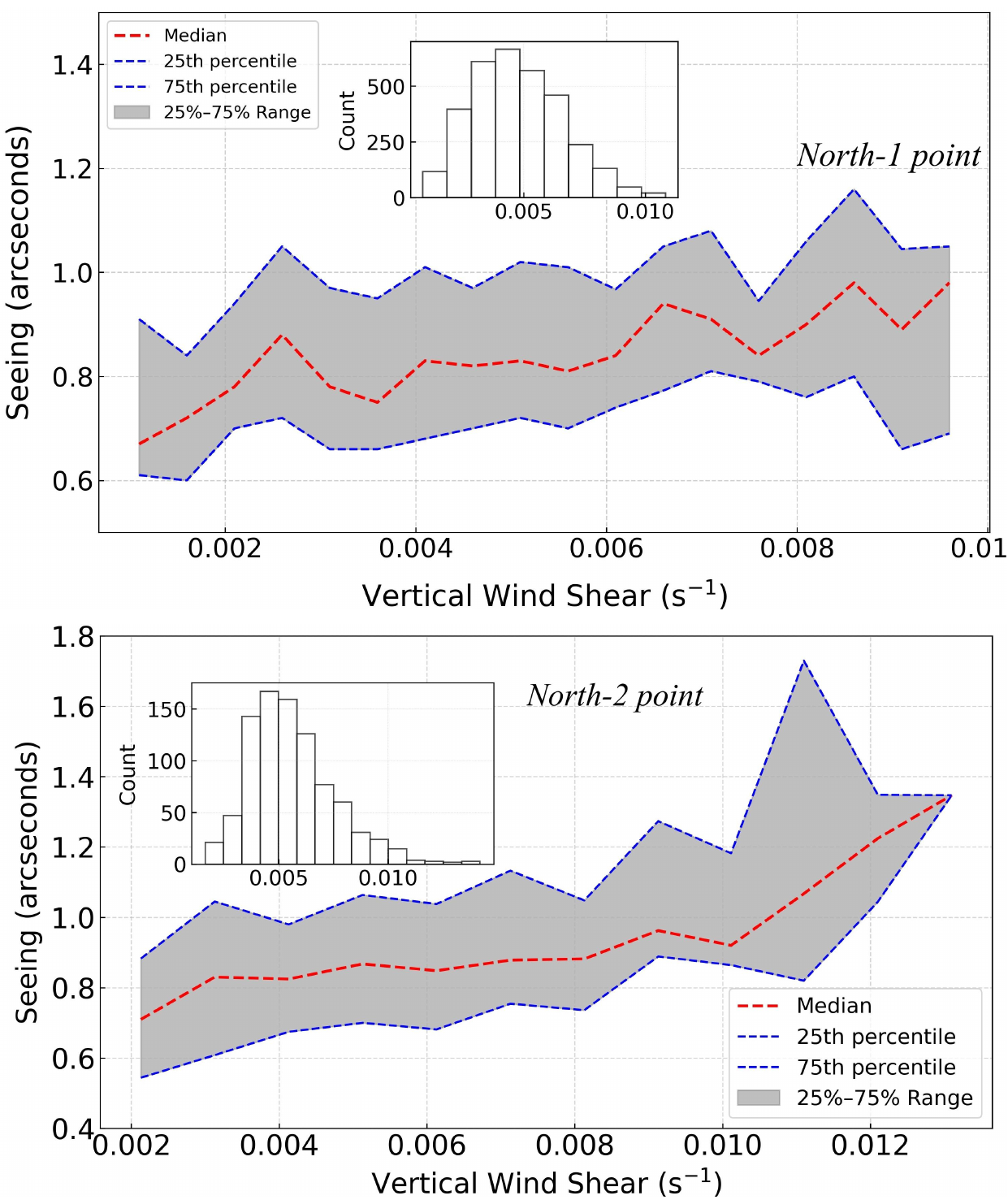}
	\caption{Relationship between seeing and average vertical wind shear at the North-1 and North-2 point. The red dashed line indicates the median seeing, and the gray shaded area represents the interquartile range (25\%--75\%).}
	\label{fig:wind_steer}
\end{figure}

Figure~\ref{fig:wind_steer} presents the statistical distribution of seeing under varying vertical wind shear conditions, with the upper panel corresponding to North-1 and the lower panel to North-2. The red dashed curve indicates the median seeing, the gray shaded region marks the 25\%–75\% interquartile range, and the inset bars show the frequency of measurements within each shear bin.
The results indicate that as vertical wind shear increases, seeing degrades at both sites, demonstrating that stronger shear corresponds to poorer seeing conditions. This effect is more pronounced at North-2, especially when shear exceeds $0.01\,\mathrm{s}^{-1}$, where the median seeing rises sharply and the distribution range widens significantly, reflecting intensified turbulence under strong shear. A smaller vertical wind shear thus helps maintain relatively stable seeing, whereas stronger shear can deteriorate observational conditions.

To assess the impact of short-term surface-layer wind fluctuations on seeing, we introduce the concept of Temporal Wind Variation, which quantifies rapid changes in wind speed over time. The variation is calculated by first determining the difference in wind speed components between consecutive time points and then applying the Euclidean norm to obtain the total change. Dividing this value by the measurement interval yields a time-normalized rate of variation, expressed as:
\begin{equation}
	\text{Temporal Wind Variation} = \frac{\sqrt{(\Delta U)^2 + (\Delta V)^2}}{\Delta t},
\end{equation}
where $\Delta U$ and $\Delta V$ represent changes in the east-west and north-south wind speed components, respectively, and $\Delta t$ is the time interval between successive measurements. The resulting unit is m/s$^{2}$.

\begin{figure}
	\centering
	\includegraphics[width=\linewidth]{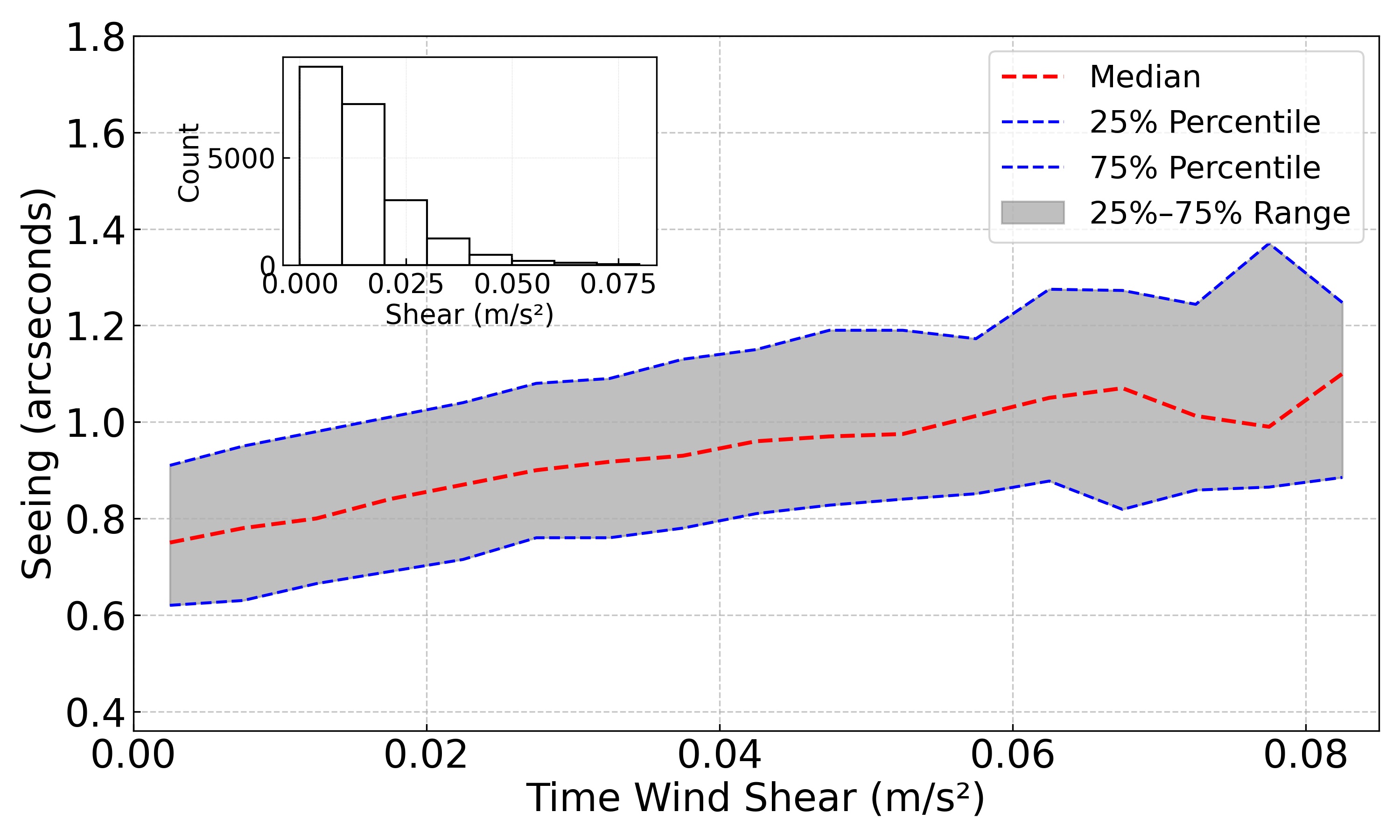}
	\caption{Relationship between seeing and average vertical time wind shear at Muztagh-ata site. The red dashed line indicates the median seeing, and the gray shaded area represents the interquartile range (25\%--75\%).}
	\label{fig:time_wind_shear_vs_median_seeing}
\end{figure}

Figure~\ref{fig:time_wind_shear_vs_median_seeing} illustrates the relationship between seeing and temporal wind variation. Overall, seeing tends to increase with stronger wind variability. When the variation remains between 0 and 0.02,m,s$^{-2}$, the median seeing stays low and stable, indicating weak turbulence and favorable observing conditions. As the variation increases, the median seeing gradually rises, and fluctuations become significantly amplified beyond 0.06,m,s$^{-2}$. These results suggest that intense short-term wind fluctuations may disrupt atmospheric stratification, enhance turbulence, and ultimately degrade optical seeing.

\section{Conclusion}
This study presents a comprehensive assessment of seeing and meteorological conditions at the North-1 and North-2 Points of the Muztagh-Ata site, based on joint monitoring campaigns conducted from 2018 to 2024. The objective was to evaluate the optical suitability of both locations as candidate sites for future large-aperture optical/infrared telescopes. The main conclusions are summarized as follows:

\begin{enumerate}
	\item \textbf{High-quality seeing statistics}: At North-1 (6\,m tower), the median seeing was 0.89\,arcseconds, with 75\% of measurements below 1.11\,arcseconds. At North-2 (10\,m tower), the median seeing was 0.78\,arcseconds, with 75\% of values below 0.97\,arcseconds. These results meet international criteria for high-quality astronomical sites, reflecting favorable atmospheric conditions at both locations.
	
	\item \textbf{Pronounced seasonal and diurnal variations}: Seeing conditions were generally more favorable in winter than in summer. Diurnally, the best seeing typically occurred shortly after sunset and before sunrise, whereas degradation near midnight was associated with intensified turbulence driven by surface radiative cooling. These trends provide practical guidance for scheduling high-precision observations.
	
	\item \textbf{Stable nighttime meteorological conditions and favorable observing availability}: The median nighttime temperature variation was 2.03$^{\circ}$C at North-1 and 2.1$^{\circ}$C at North-2. Median wind speeds ranged from 5 to 6\,m\,s$^{-1}$, with prevailing wind directions between 210$^{\circ}$ and 300$^{\circ}$, and surface pressure remained stable. The average annual Available Observing Time (AOT) was 64\%. Compared to North-2, North-1 exhibited slightly better thermal stability but lower wind speeds. These stable nighttime conditions help suppress near-surface turbulence and improve observing quality.

	\item \textbf{Significant impact of wind dynamics on seeing}: Both sites exhibited a U-shaped relationship between seeing and wind speed—moderate winds tended to suppress turbulence and improve seeing, while both low and high wind speeds were associated with degradation. In addition, strong vertical wind shear and rapid temporal fluctuations in wind speed were shown to further deteriorate seeing, highlighting the importance of wind dynamics in site selection and telescope design.
\end{enumerate}

In summary, the North-1 and North-2 Points of the Muztagh-Ata site exhibit High-quality characteristics in terms of seeing, nighttime atmospheric stability, and meteorological conditions, highlighting their strong potential to serve as world-class ground-based observatories. These results provide robust observational evidence and scientific justification for site selection efforts in the western plateau regions of China.

Looking ahead, future work will continue to focus on both North-1 and North-2, with the goal of extending the temporal coverage and frequency of seeing and turbulence monitoring. Ongoing studies will investigate the evolution of turbulence profiles and boundary layer structures, as well as their influence on optical observing conditions. Plans are also underway to deploy advanced turbulence profiling systems—such as Doppler wind lidars and radiosondes—in combination with DIMM observations, enabling a more detailed characterization of vertical turbulence distribution. These efforts will offer comprehensive theoretical and engineering support for the evaluation, design, and operational planning of next-generation large-aperture telescopes.

\section{acknowledgment}
This work is supported by the National Natural Science Foundation of China (Grant No: U2031209), Tianshan Talent Training Program (Grant No. 2023TSYCLJ0053), the Chinese Academy of Science(CAS) “Light of West China” Program(Grant No.2022 XBONXZ 014), “Sponsored by Natural Science Foundation of Xinjiang Uygur Autonomous Region” Program (Grant No. 2022D01A361), Tianshan Innovation Team Program of Xinjiang Uygur Autonomous Region(Grant No.2024D14015) , Central Guidance for Local Science and Technology Development(Grant No. ZYYD2025QY27).

\section{Data Availability}

The ERA5 reanalysis data provided by the European Centre for Medium-Range Weather Forecasts (ECMWF) are publicly accessible via the official Copernicus Climate Data Store:  
\url{https://cds.climate.copernicus.eu/cdsapp#!/dataset/10.24381/cds.bd0915c6?tab=overview}

Due to national policy restrictions, the site-specific seeing and meteorological data used in this study are not publicly available at present. Interested researchers may contact the corresponding author for more details.







\bsp	
\label{lastpage}

\end{document}